%% file: 0.tex
\documentclass[10pt]{article}
\usepackage{RR_LaBRI}
\usepackage{comment}
\usepackage{euscript}
\usepackage{amssymb}
\usepackage{amsmath}
\usepackage{amsthm}

\usepackage{diagrams}

\usepackage[all]{xy}

\input theoremsdefs.sty
\input macros.tex

\excludecomment{finalita}
\includecomment{labripreprint}

\title{$\mu$-Bicomplete Categories and Parity Games}
\author{Luigi Santocanale \\
  LaBRI - Universit\'e Bordeaux 1. \\
  {\tt santocan@labri.fr}.}

\date{}

\begin{document}

\authorLaBRI{by Luigi Santocanale} 
\dateLaBRI{September, 2002}
\LaBRIcolor
\maketitleLaBRI{RR-1281-02}

\maketitle

\input abstract.tex

\section{Introduction}

\input intro.tex

\section{Notation and Preliminaries}
\label{sec:notation}

\input preliminaries.tex

\section{$\mu$-Bicomplete Categories} 
\label{sec:mubicomplete}

\input mubicomplete.tex

\section{Parity Games as Functors}
\label{sec:paritygames}

\input parityfunctors.tex

\section{Parity Functors in the Category of Sets}
\label{sec:inset}

\input inset.tex

\section{Conclusions}
\label{sec:conclusions}

\input conclusions.tex

\newpage

\bibliographystyle{plain}
\bibliography{fics01}

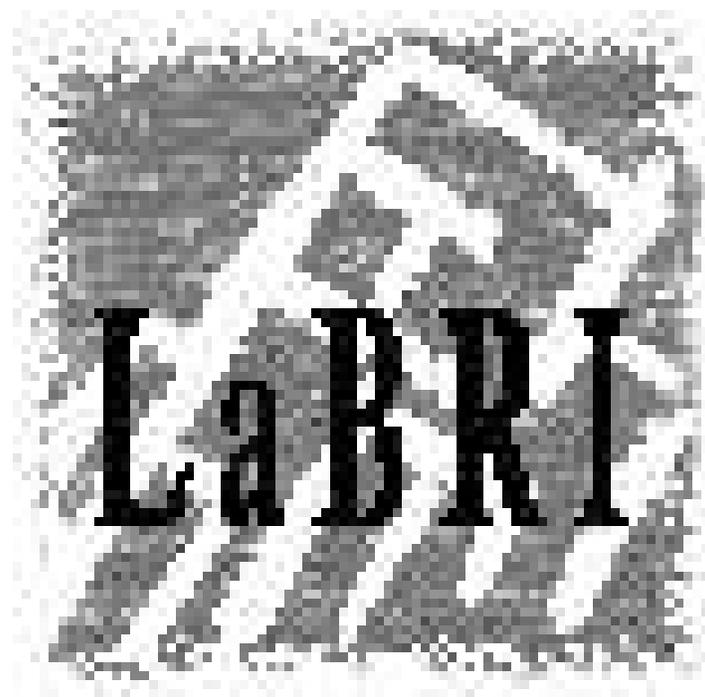
\end{document}

%% file: macros.tex
\newcommand{\keywords}[1]{\par\noindent\textbf{Keywords:} #1}

\newcommand{\length}[1]{\# #1}
\newcommand{\rise}[1]{{#1}^{\bullet}}
\newcommand{\low}[1]{{#1}_{\bullet}}
\newcommand{\tree}[1]{\tau{[#1]}}
\newcommand{\lab}[1]{\ell{[#1]}}
\newcommand{\sval}[1]{|\,#1 \,|}
\newcommand{\dgmu}[1]{{#1}^{\mu}}
\newcommand{\dgnu}[1]{{#1}^{\nu}}
\newcommand{\dom}{\partial_{0}}
\newcommand{\cod}{\partial_{1}}
\newcommand{\un}[1]{{#1}^{!}}
\newcommand{\EQ}{\mathcal{E}}
\newcommand{\hg}{h}
\newgraphescape{I}[1]{
  []*{}*+[o]{}*+\frm{}="#1"
}

\renewcommand{\lefteqn}[2][0mm]{\makebox[#1][l]{$\displaystyle{#2}$}}
\newcommand{\Cat}[1]{{\EuScript{#1}}}
\newcommand{\Sets}{{\Cat{S}\mathrm{et}}}
\newcommand{\dg}[2][]{{#2}^{\dag_{#1}}}
\newcommand{\id}{\mathtt{id}}
\newcommand{\Id}{\mathtt{id}}
\newcommand{\comp}{\cdot}
\newcommand{\compp}{\star}
\newcommand{\ltuple}{\bigl \langle}
\newcommand{\rtuple}{\bigr \rangle}
\newcommand{\prj}{{\mathtt{pr}}}
\newcommand{\inj}{{\mathtt{in}}} 
\newcommand{\lcotuple}{\bigl\{}
\newcommand{\rcotuple}{\bigr\}}
\newcommand{\iso}{\cong}
\newcommand{\nnumbers}{\mathbb{N}}
\newcommand{\mygraph}[1]{\hat{n}}
\newcommand{\set}[1]{\{1,\ldots ,#1\}}
\newcommand{\agset}[1]{\{1,\ldots ,#1,\omega\}}
\newcommand{\height}[1]{\mathrm{hg}(#1)}
\newcommand{\In}{\mathrm{In}\,}
\newcommand{\card}{\mathrm{card}\,}
\newcommand{\res}{\setminus}
\newcommand{\tiff}{\;\textrm{iff}\;}
\newcommand{\colim}[1][]{\mathrm{colim}_{#1}\,}

\newcommand{\push}[1]{+\hspace{-0.35em}{}_{#1}\hspace{0.1em}}

\newcommand{\val}[1]{\|\, #1 \,\|}
\newcommand{\bd}[1]{{\mathbf{#1}}}
\newcommand{\struct}[1]{{\if#1x\chi\else\if#1y\xi\else\if#1u\upsilon\else\if#1w\omega\else\if#1z\zeta\else\if#1a\alpha\else??\fi\fi\fi\fi\fi\fi}}
\newenvironment{system}{
    \left \{ \begin{array}{lcl}}{\end{array}\right\}
  }
\newcommand{\Land}[1][]{{\textstyle\bigwedge_{#1}}}
\newcommand{\Lor}[1][]{{\textstyle\bigvee_{\!\!#1\,}}}
\newcommand{\mydiagram}[2][6em]{\xy\xygraph{!{<#1,0cm>:<0cm,#1>::}#2}\endxy}
\newcommand{\myspecialdiagram}[3][6em]{\xy *!C=(#2,#2)\xybox{
    \xygraph{!{<#1,0cm>:<0cm,#1>::}#3}
  }\endxy
}

\newcommand{\mydiagrambot}[2][3em]{
  \begin{array}[b]{@{\hspace{0mm}}c@{\hspace{0mm}}}
    \mbox{\xy\xygraph{!{<#1,0cm>:<0cm,#1>::}#2}\endxy}
    \\[-1.3em]\hspace{2mm}
  \end{array}
  }

\newgraphescape{s}[8]{
  []
  :@{{}{}{}}[]*+{#1}="1"
  (
  :@{{}{}{}}[d(#5)r(#6)]*+{#2}="2"
  :@{{}{}{}}[d(#8)r(#7)]*+{#4}="4",
  :@{{}{}{}}[d(#8)r(#7)]*+{#3}="3"
  )
  "4"
}
\newgraphescape{S}[6]{
  []!s{#1}{#2}{#3}{#4}{#5}{0}{#6}{0}
}
\newgraphescape{a}[4]{
  [](
  "1":"2"#1,
  "1":"3"#2,
  "2":"4"#3,
  "3":"4"#4,
  )
}
\newgraphescape{A}[4]{
  []!a{^{#1}}{^{#2}}{^{#3}}{^{#4}}
}
\newcommand{\mygame}[2][3em]{
  \mbox{$\begin{array}{@{\hspace{0mm}}c@{\hspace{0mm}}}
      \xy\xygraph{!{<#1,0cm>:<0cm,#1>::}#2}\endxy
    \end{array}$}
}
\newgraphescape{p}[1]{
  []*{#1}*+[o]{}*+\frm{o}
}
\newgraphescape{P}[1]{
  []!p{#1}="#1"(
  [d(0.23)r(0.23)]*{\sigma})
}
\newgraphescape{O}[1]{
  []!p{#1}="#1"(
  [d(0.23)r(0.23)]*{\pi})
}
\newgraphescape{E}[1]{
  []!p{#1}="#1"
}
\newgraphescape{g}[1]{
  []*{#1}*+[o]{}*+\frm{}
}
\newgraphescape{L}[2]{
  []="Frame2"
  ?="Frame1"
  !{{{"Frame1"."Frame2"}*++\frm{--}}="Frame3"}
  (
  []!{"Frame3"!DL}[u(0.1)l(0.2)]*{#1},
  []!{"Frame3"!UR}[d(0.1)r(0.2)]*[r]{#2}
  )
}

%% file: abstract.tex
\begin{abstract}
For an arbitrary category, we consider the least class of functors
containing the projections and closed under finite products, finite
coproducts, parameterized initial algebras and parameterized final
coalgebras, i.e.  the class of functors that are definable by
$\mu$-terms. We call the category $\mu$-bicomplete if every $\mu$-term
defines a functor. We provide concrete examples of such categories and
explicitly characterize this class of functors for the category of
sets and functions.  This goal is achieved through parity games: we
associate to each game an algebraic expression and turn the game into
a term of a categorical theory.  We show that $\mu$-terms and parity
games are equivalent, meaning that they define the same property of
being $\mu$-bicomplete.  Finally, the interpretation of a parity game
in the category of sets is shown to be the set of deterministic
winning strategies for a chosen player.

\vspace{1em}

\keywords{parity games, bicomplete categories, initial algebras, final coalgebras, inductive and coinductive types.}
\end{abstract}

%% file: intro.tex
Several set-theoretic structures of relevance to computer science can
be described either by using the language of initial algebras or by
using the language of final coalgebras.  For example, sets of finite
trees, the set of terms over a signature and, more in general,
inductively defined sets are initial algebras of some functor and this
property characterizes these sets up to canonical isomorphism.
Similarly, sets of trees with possibly infinite branches, sets of
objects which are canonical solutions of systems of equations,
coinductively defined sets or coinductive types, can be characterized
up to isomorphism by the property of being final coalgebras of some
functor.  Thus the study of initial algebras in connection with data
structures is a well developed subject and dates back at least twenty
five years \cite{MR82d:18004,lehmannsmyth}. The interest in final
coalgebras is more recent, but is nowadays grown up to a well
established discipline \cite{MR89j:03039,barr,MR1791953}.  Despite the
existence of programming languages and proof assistants that implement
both inductive and coinductive types \cite{CoFu92,gimenez}, it appears
to us that initial algebras and final coalgebras are not often studied
simultaneously in the literature. Thus we are led to the problem of
understanding what kind of structures arise from both initial algebras
and final coalgebras and whether these structures are of any interest
in computer science.

This paper is therefore meant to be an introduction to the theory of
categories having the following completeness properties: they have
finite \emph{sums}, finite \emph{products} and all the \emph{initial
  algebras} and \emph{final coalgebras} of the functors which can be
constructed out of these four operations, using projection functors as
building blocks.  These categories generalize $\mu$-lattices
\cite{San99:Fml,BRICS-RS-00-29} and $\mu$-algebras \cite{MR87e:03079}
on one side, on the other side they partially generalize bicomplete
categories \cite{MR96k:18004a}.  For this reason we call them
\emph{$\mu$-bicomplete}.

A first concern is to show that many common categories are
$\mu$-bicomplete. It is well known that an accessible unary
endofunctor of a locally presentable category has both an initial
algebra and a final coalgebra. We adapt ideas existing in the
literature to prove that locally presentable categories are
$\mu$-bicomplete.  Among the locally presentable categories is the
category of sets and functions, which is therefore $\mu$-bicomplete.

 As it is often the case for
existence theorems, the mere knowledge that a category is
$\mu$-bicomplete is unsatisfactory. The principal goal of this paper
is that of giving an explicit description of the functors on the
category of sets that arise out of those four operations, i.e. of the
functors that are definable by $\mu$-terms.   We achieve this goal by translating the algebraic
language of $\mu$-bicomplete categories into the combinatorial
language of \emph{parity games}, cf. \cite[\S 4]{AN01}.  These games
are a standard tool in the theory of automata recognizing infinite
objects \cite{thomas}. A central notion in this theory is that of an
acceptance condition, essentially a method for specifying a set of
infinite paths in a graph. The acceptance condition by which the set
of infinite winning plays in a parity game is defined was introduced
in \cite{MR827531} to construct automata in normal form.  Thus it
should not come as a surprise that several combinatorial problems of
the theory can be reduced to the problem of finding winning strategies
in a parity game.  For example, the properties of transition systems
that are definable by alternating fixed point expressions can be
checked using algorithms designed and proved correct by means of
game-theoretic ideas and analogies \cite{emersonjutla}.

Generalizing ideas that relate the theory of two persons games with
the theory of bicomplete categories \cite{MR96k:18004a,MR1797101}, we
show that it is possible to endow parity games with an algebraic
meaning, so that they can be considered to be terms of a categorical
theory.  We show then the equivalence of this meaning to the one of
$\mu$-terms defining $\mu$-bicomplete categories.  On the
combinatorial side, parity games can be considered as recognizers of
infinite objects in a natural way: they recognize the set of
deterministic winning strategies for a chosen player. The two
different meanings of parity games, the algebraic one and the
combinatorial one, are then shown to coincide if the category of sets
and functions is being considered.  This leads to the characterization
of functors denoted by $\mu$-terms in the category of sets: a
$\mu$-term is translated into a parity game and its denotation in this
category is the set of deterministic winning strategies for the chosen
player.

This result
supports the claim that the algebra of parity games is the one of
$\mu$-bicomplete categories and that the combinatorics of
$\mu$-bicomplete categories is the one of parity games, a claim which
is meant to emphasize two possible directions of research. One goes
from the algebra to the combinatorics: for example, this work should
provide a starting point for elaborating game semantics of programming
languages that implement both inductive and coinductive types.  The
other direction goes from the combinatorics to the algebra: we are
proposing an alternative algebraic interpretation of the alternation
between ``finitely many times'' and ``infinitely often'' which occurs
so often in the theory of automata recognizing infinite objects. The
alternation is usually analyzed by means of complete lattices, of
ordinals and of approximants of least fixed points that occur nested
within greatest fixed points.  Our interpretation requires induction
and coinduction, that is, initial algebras and final coalgebras of
functors which are natural generalizations of least and greatest fixed
points.  A particular motivation for developing this work has been the
possibility of describing transformations of winning strategies by
means of arrows definable in every $\mu$-bicomplete category.  We have
reported partial results on the structure of arrows of
$\mu$-bicomplete categories in \cite{sanlccp}.

On several occasions categories with similar completeness properties
have been proposed. For example, in \cite{MR93g:18005} a category
$\Cat{C}$ is defined to be algebraically complete if every unary
endofunctor has an initial algebra.  This requirement appears to be
too strong, since the only complete categories with this property turn
out to be the complete quasi-orders. It is possible to relax the
requirement and ask only a given class of functors of the form
$F:\prod_{i \in I}\Cat{C} \rTo \Cat{C}$ to be closed under
parameterized initial algebras.  This approach is the one proposed in
\cite{MR98j:18007} and leads to define $2$-iteration theories
\cite{MR2001m:18006}.  This is also our approach with the proviso that
we are interested in a specific class which is required to be closed
under parameterized final coalgebras as well. In \cite{MR93g:18005}
final coalgebras are considered too, but they are flattened into
initial algebras: an algebraically compact category is defined there
to be an algebraically complete category such that the inverse of
every initial algebra is also a final coalgebra of the same functor.

In these and other contexts the equational properties of categorical
fixed points have been studied. Our aim here is to see these
properties at work.  Our starting point will be the Beki\v{c}
property. In its simplest form it is an inductive method for showing
that a system of equations admits a unique solution: a sufficient
condition for the system of equations
$$
\begin{system}
  x & = & f(x,y) \\
  y & = & g(x,y)
\end{system}
$$
\vfill\eject\noindent
to admit a unique solution is that
the system
$$
\begin{system}
  x & = & f(x,y)
\end{system}
$$
admits a unique solution $x = \dg{f}(y)$ for each choice of $y$ and
that either of the two equivalent systems
$$
\begin{system}
  x & = & \dg{f}(y) \\
  y & = & g(x,y)
\end{system}
\;\;\;\;
\begin{system}
  y & = & g(\dg{f}(y),y)
\end{system}
$$
admits a unique solution. The analogy of the Beki\v{c} property
with Gaussian elimination has been pointed in
\cite{simpsonplotkin,AN01}. The equivalence between the last two
systems is moreover the root of our algebraic interpretation of parity
games.  We shall state a categorical version of the Beki\v{c}
property; the property stated above is recovered when considering that
a set is a poset with a discrete order, and that a poset is a category
with at most one arrow between any two objects: an initial algebra of
an endofunctor of a discrete category is nothing else but a unique
fixed point.  The Beki\v{c} property allows to prove the equivalence
between $\mu$-terms and systems of functorial equations. These seem to
be better suited than $\mu$-terms for an analysis that emphasizes the
operational aspects.  \vspace{2mm}

The paper is structured as follows. In section \ref{sec:notation} we
explain the notation, introduce the principal concepts and state the
Beki\v{c} property.  In section \ref{sec:mubicomplete} we define
categorical $\mu$-terms and $\mu$-bicomplete categories. We prove that
locally presentable categories are $\mu$-bicomplete. In section
\ref{sec:paritygames} we define parity games and their algebraic
interpretation. We show that it is possible to interpret every parity
game on a category if and only if the category is $\mu$-bicomplete, by
giving a translation of parity games into $\mu$-terms and vice-versa.
In section \ref{sec:inset} we prove that the algebraic interpretation
of a parity game in the category of sets is the set of winning
strategies for a chosen player.  We add some examples and applications
of the theory so far developed.  Finally, in section
\ref{sec:conclusions}, we add concluding remarks.

%% file: preliminaries.tex
\subsection{Notation} 
We will use different notations for the categorical composition.
Given two arrows $f : A \rTo B$ and $g : B \rTo C$ of a category
$\Cat{C}$, we  use the notation $f \comp g : A \rTo C$ for their
composition. However, when dealing with functors $F:\Cat{C} \rTo
\Cat{D}$ and $G : \Cat{D} \rTo \Cat{E}$, we  prefer the notation
$G \circ F$, or simply $GF$. In a similar way, if $f: A \rTo B$ is a
set-theoretic function, we  use $f(a)$, $fa$ and $f_{a}$ for
evaluation at $a \in A$. 

\begin{labripreprint}
  If $w : I \rTo \Cat{W}$ is a functor, we use the notation
  $(\inj_{i}: w_{i} \rTo\colim w)_{i \in I}$ for a chosen colimiting
  cocone. If $(f_{i}:w_{i} \rTo c)_{i \in I}$ is another cocone, then
  we let $\lcotuple f_{i}\rcotuple_{i \in I}:\colim w \rTo c$ be the
  unique arrow such that $\inj_{i} \comp \lcotuple f_{i}\rcotuple_{i
    \in I} = f_{i}$, for all objects $i$ of $I$. We use analogous
  notations $\prj_{i}, \lim$, and $\ltuple f_{i} \rtuple_{i \in I}$
  for chosen limiting cones.
\end{labripreprint}

We use the symbols $\dom,\cod$ for the domain and codomain functions
of graphs and categories. Given a graph $S = \langle \dom,\cod: M\rTo
P\rangle$, we write $m:p \rightarrow q$ to mean that $m \in M$, $\dom
m = p$ and $\cod m = q$. The free category over the graph $S$ is
described as follows: its set of objects is $P$ and an arrow from $p$
to $q$ is a sequence of transitions $\gamma = \lcotuple m_{i}
\rcotuple_{i =1,\ldots ,n}$ such that $\dom m_{1} = p$, $\dom m_{i +1}
= \cod m_{i}$, $i =1,\ldots n -1$ and $\cod m_{n} = q$, that is, it is
a path in $S$ from $p$ to $q$; we say in this case that $n =
\length{\gamma}$ is the length of $\gamma$.  Given two paths $\delta$
from $p$ to $q$ and $\gamma$ from $q$ to $r$ in $S$, we use the
notation $\delta \compp\gamma$ for their composition.  The identity of
a vertex $p$ is the path $1_{p}$ from $p$ to $p$ having null length. A
path $\delta$ is a prefix of $\gamma$ if there exists a path $\gamma'$
such that $\gamma = \delta \compp \gamma'$.

A morphism of graphs $\Phi: \langle P_{1},M_{1},\dom,\cod\rangle \rTo
\langle P_{2},M_{2},\dom,\cod\rangle$ is a pair of functions $\Phi :
P_{1} \rTo P_{2}, \Phi: M_{1} \rTo M_{2}$ such that
$\partial_{i}\Phi(m) = \Phi(\partial_{i}m)$, for $i =0,1$ and $m \in
M_{1}$. We can describe a path in $S$ as a morphism of graphs $\gamma:
\hat{n} \rTo S$, where $\hat{n}$ is the graph $0 \rightarrow 1
\rightarrow \ldots \rightarrow n$.  An infinite path in $S$ is a
morphism of graphs $\gamma : \hat{\omega} \rTo S$, where
$\hat{\omega}$ is the graph $0 \rightarrow 1 \rightarrow \ldots
\rightarrow n \rightarrow \ldots $. If $\delta$ is a finite path from
$p$ to $q$ and $\gamma$ is an infinite path such that $\gamma_{0} =
q$, then we write $\delta\compp \gamma$ for the resulting infinite
path. A morphism of graphs $\Phi: S_{1} \rTo S_{2}$ induces a functor
between the respective free categories, which we will denote by the
same letter $\Phi$. Observe that if $\gamma$ is a path in $S_{1}$,
then $\Phi(\gamma)$ is the morphism of graphs $\gamma \comp \Phi$,
thus we extend the same notation to infinite paths, letting in this
case $\Phi(\gamma) = \gamma \comp \Phi$.

\subsection{Initial Algebras of Functors}

Let $\Cat{C}$ be a category and $F:\Cat{C} \rTo \Cat{C}$ be an
endofunctor, an \emph{$F$-algebra} is a pair $(c,\gamma)$, where $c$
is an object of $\Cat{C}$ and $\gamma: Fc \rTo c$ is an arrow of
$\Cat{C}$. A morphism of $F$-algebras $f:(c,\gamma) \rTo (d,\delta)$
is an arrow $f:c \rTo d$ of $\Cat{C}$ such that $\gamma \comp f = Ff
\comp \delta$. $F$-algebras and their morphisms form a category
$\Cat{C}^{F}$ and we define an initial $F$-algebra to be an initial
object in this category. More explicitly, an $F$-algebra
$(\bd{x},\struct{x})$ is initial if for each $F$-algebra $(c,\gamma)$
there exists a unique arrow $f: \bd{x} \rTo c$ such that $\struct{x}
\comp f = Ff \comp \gamma$.  We remark that if an $F$-algebra
$(\bd{x},\struct{x})$ is initial, then the arrow $\struct{x}$ is
invertible \cite{lambek68}.

$F$-coalgebras and their morphisms are
defined dually and form a category $\Cat{C}_{F}$. We recall that a
coalgebra $\struct{y}:\bd{y}\rTo F\bd{y}$ is final if for each
coalgebra $\gamma: c \rTo Fc$ there exists a unique arrow $g: c \rTo
\bd{y}$ such that $g\comp \struct{y} = \gamma \comp Fg$.  

If $F: \Cat{C}\times \Cat{D} \rTo \Cat{C}$ is such that for every
object $d$ of $\Cat{D}$ there exists an initial algebra
$(\dgmu{F}(d),\struct{x}_{d})$ of the functor $F(-,d)$, then there
exists a unique way to turn the collection of objects $\dgmu{F}(d)$
into a functor so that $\struct{x}_{d}: F(\dgmu{F}(d),d) \rTo
\dgmu{F}(d)$ is a natural isomorphism: for $f: d \rTo d'$,
$\dgmu{F}(f)$ is the unique $F(-,d)$-algebra morphism from the initial
one $(\dgmu{F}(d), \struct{x}_{d})$ to $(\dgmu{F}(d'),
F(\dgmu{F}(d'),f)\comp\struct{x}_{d'})$. We call the arising functor
$\dgmu{F}: \Cat{D} \rTo \Cat{C}$ a \emph{parameterized initial
  algebra} of $F$.  A parameterized final coalgebra $\dgnu{F}$ of $F$
is defined similarly.

\subsection{The Beki\v{c} Property}
\label{sec:bekic}

We state here the Beki\v{c} property for initial algebras of functors,
a proof of which is found in \cite[\S 4.2]{lehmannsmyth}.  This
property will be a major tool in the proofs that follow.
\begin{labripreprint}
  Thus we add here a proof, with the aim of making this report self
  contained.
\end{labripreprint}
\begin{prpstn}
  \label{prop:bekic}
  Consider two functors $F : \Cat{C} \times \Cat{D} \rTo \Cat{C}$, $G
  : \Cat{C} \times \Cat{D} \rTo \ \Cat{D}$, and for each object $d$ of
  $\Cat{D}$ let $(\dgmu{F}(d),\struct{x}_{d})$ be an initial
  $F(-,d)$-algebra.  Suppose moreover that there exists an initial
  algebra
  \begin{align*}
    \struct{y} &: \dgmu{F}(\bd{z}) \rTo  \bd{y}
    &
    \struct{z} & :  G(\bd{y},\bd{z})  \rTo^{}
    \bd{z}
  \end{align*}
  of the functor $\ltuple \dgmu{F} \circ \prj_{\Cat{D}}, G \rtuple:
    \Cat{C}\times \Cat{D} \rTo \Cat{C}\times \Cat{D}$.  Then the pair
  \begin{align*}
    \struct{x}_{\bd{z}}& : 
    F(\dgmu{F}(\bd{z}),\bd{z})
    \rTo \dgmu{F}(\bd{z}) &
    G(\struct{y},\bd{z})\comp \struct{z} & 
    : G(\dgmu{F}(\bd{z}),\bd{z})\rTo \bd{z}
  \end{align*}
  is an initial algebra of the functor $\ltuple F, G \rtuple : \Cat{C}
  \times \Cat{D} \rTo \Cat{C} \times \Cat{D}$.
\end{prpstn}
\begin{labripreprint}
  \begin{proof}
    Given an $\ltuple F, G \rtuple$-algebra $(\langle c,d\rangle,
    \langle\gamma,\delta\rangle)$, let $\un{\gamma}$ be the unique
    arrow such that $\struct{x}_{d} \comp \un{\gamma} =
    F(\un{\gamma},d) \comp \gamma$.  Then the pair $(\langle
    c,d\rangle,\langle\un{\gamma},\delta\rangle)$ is an $\langle
    \dgmu{F} \circ \prj_{\Cat{D}},G\rangle$-algebra and therefore we can
    find a pair of arrows $f: \bd{y} \rTo c$ and $g: \bd{z} \rTo d$
    satisfying the equations
    \begin{eqnarray*}
      \struct{y}\comp f
      & = & \dgmu{F}(g) \comp \un{\gamma} \\
      \struct{z} \comp g
      & = & G(f,g) \comp \delta\
    \end{eqnarray*}
    and unique with this property.  By inspecting the two diagrams
    $$
    \mydiagram[6em]{
      [](!S{F(\dgmu{F}(\bd{z}),\bd{z})}{\dgmu{F}(\bd{z})}
      {F(\dgmu{F}(d),d)}{\dgmu{F}(d)}
      {1}{1.7}
      !A
      {\struct{x}_{\bd{z}}}{F(\dgmu{F}(g),g)}
      {\dgmu{F}(g)}{\struct{x}_{d}}
      )
      "3"
      (
      !S{F(\dgmu{F}(d),d)}{\dgmu{F}(d)}
      {F(c,d)}{c}
      {1}{1.5}
      !A
      {}{F(\un{\gamma},d)}
      {\un{\gamma}}
      {\gamma}
      )
    }
    $$
    $$
    \mydiagram[6em]{
      [](
      !S
      {G(\dgmu{F}(\bd{z}),\bd{z})}
      {G(\dgmu{F}(d),\bd{z})}{G(\bd{y},\bd{z})}
      {G(c,d)}
      {1}{1.7},
      !A
      {G(\dgmu{F}(g),\bd{z})}{G(\struct{y},\bd{z})}
      {G(\un{\gamma},g)}
      {G(f,g)}
      )
      "3"
      (
      !S{G(\bd{y},\bd{z})}
      {G(c,d)}{\bd{z}}{d}
      {1}{1.5},
      !A
      {}{\struct{z}}
      {\delta}
      {g}
      )   
    }
    $$
    we deduce that the relations 
    \begin{eqnarray*} 
      \struct{x}_{\bd{z}} \comp \,( \,\dgmu{F}(g) \comp \un{\gamma}\,)\,
      & = & F(\dgmu{F}(g)\comp \un{\gamma},g)\comp \gamma \\ 
      (\,G(\struct{y},\bd{z})\comp \struct{z}\,)\,\comp g
      & = & G(\dgmu{F}(g) \comp \un{\gamma},g)  \comp \delta 
    \end{eqnarray*}
    hold.   On the other hand, suppose that
    \begin{eqnarray*} 
      \struct{x}_{\bd{z}} \comp h
      & = & F(h,k)\comp \gamma \\ 
      (\,G(\struct{y},\bd{z})\comp \struct{z}\,)\,\comp k
      & = & G(h,k)  \comp \delta 
    \end{eqnarray*}
    hold, then $h$ is the unique morphism of $F(-,\bd{z})$-algebras
    from the initial one to $(c,F(c,k) \comp \gamma)$.  If we take $k$
    in place of $g$ in the first diagram, we obtain that $\dgmu{F}(k)
    \comp \un{\gamma}$ is also a morphism of $F(-,\bd{z})$-algebras
    from the initial one to $(c,F(c,k) \comp \gamma)$ and therefore
    \begin{eqnarray*}
      h
      & = & \dgmu{F}(k) \comp \un{\gamma}\,.
    \end{eqnarray*}
    The two equations
    \begin{eqnarray*}
      \struct{y} \comp (\struct{y}^{-1}\comp h) 
      & = & \dgmu{F}(k) \comp \un{\gamma} \\
      \struct{z} \comp k & = & 
      G(\struct{y}^{-1}\comp h,k ) \comp \delta
    \end{eqnarray*}
    hold and show that $\struct{y}^{-1}\comp h = f$ and $k = g$, since
    a pair satisfying the above relations is unique.
  \end{proof}
\end{labripreprint}

The following proposition is needed to obtain the Beki\v{c} lemma in
its usual form, see \cite{lehmannsmyth} or the pairing identity in
\cite[\S 5.3.9]{MR95g:68065}.
\begin{prpstn}
  \label{prop:bekic2}
  \label{prop:bekic3}
  Consider two functors $F: \Cat{D} \rTo \Cat{C}$, $G :
  \Cat{C}\times \Cat{D} \rTo \Cat{D}$, and let
  $(\bd{y},\struct{y})$ be an initial algebra of the functor
  $G(F-,-):\Cat{D} \rTo \Cat{D}$. Then the pair
  \begin{align*}
    \id_{F(\bd{y})}
    & :  F(\bd{y})  \rTo F(\bd{y})
    &
    \struct{y} & :  G(F(\bd{y}),\bd{y})
    \rTo \bd{y} 
  \end{align*}
  is an initial algebra of the functor $\langle F \circ\prj_{\Cat{D}
  }, G\rangle: \Cat{C} \times \Cat{D} \rTo \Cat{C} \times \Cat{D}$.
  Conversely, if
  \begin{align*}
    \struct{x} & : F(\bd{y}) \rTo \bd{x} 
    & 
    \struct{y} & :
    G(\bd{x},\bd{y}) \rTo \bd{y}
  \end{align*}
  is an initial algebra of the functor $\langle F \circ\prj_{\Cat{D}},
  G\rangle$, then
  \begin{align*}
    G(\struct{x},\bd{y})\comp \struct{y}
    & :  G(F(\bd{y}),\bd{y})
    \rTo \bd{y}
  \end{align*}
  is an initial algebra of the functor $G(F-,-):\Cat{D} \rTo \Cat{D}$.
\end{prpstn}
\begin{labripreprint}
\begin{proof}
 Let $(\langle c,d \rangle,\langle \gamma,\delta
  \rangle)$ be an $\langle F\circ\prj_{\Cat{D}},G\rangle$-algebra and
  observe that if the relations
  \begin{eqnarray*} 
    \id_{F(\bd{y})} \comp f & = & Fg \comp \gamma \\
    \struct{y} \comp g & = & G(f,g) \comp \delta 
  \end{eqnarray*}
  hold, then the relation
  \begin{eqnarray*}
    \struct{y} \comp g & = & G(Fg \comp \gamma,g) \comp \delta \\
    & = & G(Fg,g) \comp G(\gamma,d) \comp \delta
  \end{eqnarray*}
  shows that $g$ is the unique morphism of $G(F-,-)$-algebras from the
  initial one to $(d,G(\gamma,d) \comp \delta)$. Since moreover $f =
  Fg \comp \gamma$, then a pair $(f,g)$ with these properties is
  uniquely determined. Hence let $g$ be this unique morphism and let
  $f = Fg \comp \gamma$, then
  \begin{eqnarray*}
    \struct{y} \comp g & = & 
    G(Fg,g) \comp G(\gamma,d) \comp \delta \\
    & = & 
    G(Fg \comp\gamma,g) \comp \delta \\
    & = & 
    G(f,g) \comp \delta 
  \end{eqnarray*}
  so that the pair $(f,g)$ is the required morphism of $\langle
  F\circ\prj_{\Cat{D}},G\rangle$-algebras.

  For the converse, let $\delta : G(Fd,d) \rTo d$ be a
  $G(F-,-)$-algebra and suppose that we can find an arrow $g$ such
  that
  \begin{eqnarray*}
    G(\struct{x},\bd{y})\comp \struct{y} \comp g
    & = & G(Fg, g) \comp \delta\,.
  \end{eqnarray*}
  Let $f = \struct{x}^{-1} \comp Fg$, then the two  relations 
  \begin{eqnarray*}
    \struct{x} \comp f & = & Fg \\
    \struct{y} \comp g & = & G(f,g) \comp \delta
  \end{eqnarray*}
  hold, so that the pair $(f,g)$ is a morphism of $\langle F \circ
  \prj_{\Cat{D}},G \rangle$-algebras from an initial one to $(\langle
  Fd,d\rangle,\langle\id_{Fd}, \delta\rangle)$. It follows that such a
  $g$ is uniquely determined. On the other hand, if  $(f,g)$ is 
  this unique morphism of $\langle F \circ
  \prj_{\Cat{D}},G \rangle$-algebras, then
  \begin{eqnarray*}
    G(\struct{x},\bd{y}) \comp \struct{y} \comp 
    g & = & G(\struct{x},\bd{y}) \comp G(f, g) \comp \delta \\
    & = & G(\struct{x}\comp f, g) \comp \delta \\
    & = & G(Fg, g) \comp \delta\,, 
  \end{eqnarray*}
  i.e. $g$ is the required $G(F-,-)$-algebra morphism.
\end{proof}  
\end{labripreprint}
The reader will have no difficulties to adapt the statements
\begin{labripreprint}
  and the proofs
\end{labripreprint}
of propositions \ref{prop:bekic} and
\ref{prop:bekic3} to construct a \emph{parameterized} initial algebra
of a functor $\ltuple F, G \rtuple : 
\Cat{C} \times \Cat{D}\times \Cat{E} \rTo \Cat{C} \times
\Cat{D}$, given a parameterized initial algebra $\dgmu{F}:
\Cat{D}\times \Cat{E} \rTo \Cat{C}$ of the functor $F : \Cat{C}\times
\Cat{D}\times \Cat{E} \rTo \Cat{C}$ and a parameterized initial
algebra of the functor $G \circ \langle \dgmu{F}, \id_{\Cat{D}\times
  \Cat{E}}\rangle : \Cat{D}\times \Cat{E} \rTo \Cat{D}$.

%% file: mubicomplete.tex
We define $\mu$-bicomplete categories by mimicking the definition of
$\mu$-algebras \cite{MR87e:03079} at the level of categories:
$\mu$-terms are defined and an algebra is a $\mu$-algebra if it is
possible to interpret all the $\mu$-terms as expected.  In a
categorical context a $\mu$-term is to be interpreted as a functor,
which generalizes the usual interpretation of a $\mu$-term as an order
preserving function.

In the following definition we explicitly keep track of free variables
in a $\mu$-term by means of a context $X$: this is simply a finite set
(of variables). Later we shall use the notation $\Cat{C}^{X}$ to
denote the $X$-fold product of a category $\Cat{C}$ with itself.

\begin{dfntn}
  The set $\mu\mathcal{T}(X)$ of $\mu$-terms over a context $X$ is
  defined as follows:
  \begin{enumerate} 
  \item For each pair $(X,x)$, where $X$ is a finite set and $x \in
    X$, $x \in \mu\mathcal{T}(X)$.
  \item If $I$ is a finite set and $s : I \rTo \mu\mathcal{T}(X)$,
    then $\Land[I] s, \Lor[I] s \in  \mu\mathcal{T}(X)$.
  \item If $s \in \mu\mathcal{T}(X)$ and $x \in X$, then $\mu_{x}.s,
    \nu_{x}.s \in \mu\mathcal{T}(X \setminus \{ x \})$.
  \end{enumerate}
\end{dfntn}

\begin{dfntn}
  Let $\Cat{C}$ be a category with finite products and finite
  coproducts. We define a partial interpretation of $\mu$-terms $s \in
  \mu\mathcal{T}(X)$ over a context $X$ as functors of the form
  $\val{s}: \Cat{C}^{X} \rTo \Cat{C}$.
  \begin{enumerate}
  \item For $x \in X$, we let $\val{x} = \prj_{x}:
    \Cat{C}^{X} \rTo \Cat{C}$.
  \item We let $\val{\Land[I] s} = \prod_{i \in I} \val{s_{i}}$ and
    $\val{\Lor[I] s} = \coprod_{i \in I} \val{s_{i}}$, given that all
    the $\val{s_{i}}$ are defined.
  \item We let $\val{\mu_{x}.s}$ be the parameterized initial algebra
    of
    $$
    \val{s}: \Cat{C}^{\{x\}} \times \Cat{C}^{X \setminus \{x\}}
    \rTo \Cat{C}\,,
    $$
    given that $\val{s}$ is defined. Similarly we let
    $\val{\nu_{x}.s}$ be the parameterized final coalgebra of
    $\val{s}$. If $\val{s}$ is not defined or if the desired initial
    algebras (final coalgebras) do not exist, then we leave
    $\val{\mu_{x}.s}$ ($\val{\nu_{x}.s}$) undefined.
  \end{enumerate}
\end{dfntn}

\begin{dfntn}
  A category with finite products and finite coproducts $\Cat{C}$ is
  said to be \emph{$\mu$-bicomplete} if for each finite set of
  variables $X$ and $\mu$-term $s \in \mu\mathcal{T}(X)$ the
  interpretation $\val{s}$ is defined.
\end{dfntn}

An alternative point of view emphasizes the class of functors which
are definable by means of $\mu$-terms in a $\mu$-bicomplete category.
Thus we are lead to give the following definition.

\begin{dfntn}
  We say that a functor $F : \Cat{C}^{X} \rTo \Cat{C}^{Y}$ is a
  \emph{$\mu$-functor} if there exist a collection of $\mu$-terms $\{
  s_{y} \in \mu\mathcal{T}(X) \}_{y \in Y}$
  and a natural isomorphism $F \iso \ltuple \val{s_{y}} \rtuple_{y \in
    Y}$.
\end{dfntn}

\begin{prpstn}
  $\mu$-functors are closed under composition.
\end{prpstn}
\begin{proof}
  Let $s \in \mu\mathcal{T}(X)$ and $z \not\in X$, we first define
  $s^{z} \in \mu\mathcal{T}(\{z \} \cup X)$ with the property that
  $\val{s^{z}} = \val{s}\circ \prj_{X}$, by induction on the structure
  of $s$.  We let $x^{z} = x$, $(\Land[I] s)^{z} = \Land[I] s^{z}$ and
  $(\Lor[I] s)^{z} = \Lor[I] s^{z}$, where $(s^{z})_{i} =
  (s_{i})^{z}$, $(\mu_{x}.s)^{z} = \mu_{x}.(s^{z})$ and
  $(\nu_{x}.s)^{z} = \nu_{x}.(s^{z})$. In the last two cases we have
  supposed that the variable $x \not\in \{z\} \cup X$, otherwise we
  can rename $x$ in $s$ to a variable $x' \not\in \{z\}\cup X $ and
  obtain a $\mu$-term $t \in \mu\mathcal{T}(\{x'\} \cup X)$ such that
  $\val{t} = \val{s}$ and $\val{\mu_{x'}.t} = \val{\mu_{x}.s}$ and
  then we can define $(\mu_{x}.s)^{z} = \mu_{x'}.(t^{z})$.
  
  Let $s : Y \rTo \mu\mathcal{T}(X)$ be a collection of $\mu$-terms
  and let $t \in \mu\mathcal{T}(Y)$. We define now a $\mu$-term $t[s]
  \in \mu\mathcal{T}(X)$ with the property that $\val{t[s]} \iso
  \val{t} \circ \ltuple \val{s_{y}}\rtuple_{y \in Y}$, by induction of
  the structure of $t$.  We let $y[s] = s_{y}$, $(\Land[I] t)[s] =
  \Land[I] (t[s])$, $(\Lor[I] t)[s] = \Lor[I] (t[s])$ where $t[s]_{i}
  = t_{i}[s]$.  Eventually, we let $(\mu_{x}.t)[s] =
  \mu_{x}.(t[x,s^{x}])$ and $(\nu_{x}.t)[s] 
    = \nu_{x}.(t[x,s^{x}])$, where $(x,s^{x}): \{x\}\cup Y \rTo
  \mu\mathcal{T}(\{x \} \cup X)$ is such that $(x,s^{x})_{y} =
  s_{y}^{x}$ if $y \in Y$ and $(x,s^{x})_{x} = x$.  We have supposed
  again and without loss of generality that $x \not\in X$. The desired
  statement follows.
\end{proof}

Let $Y$ be a set of variables and suppose that it is the disjoint
union of $X$ and $Z$.  We can extend every collection $s : Z \rTo
\mu\mathcal{T}(X)$ indexed by $Z$ to a collection $s' : Y \rTo
\mu\mathcal{T}(X)$ by letting $s'_{y} = s_{y}$ if $y \in Z$ and
$s'_{y} = y$ if $y \in X$. Thus if $t \in \mu\mathcal{T}(Y)$ then we
let
\begin{eqnarray*}
  t[s_{z}/z]_{z\in Z} & = & t[s']
\end{eqnarray*}
where $t[s']$ has been defined in the proof of the above proposition.
We observe that the interpretation of $\val{t[s_{z}/z]_{z\in Z}}$ is
$$
\Cat{C}^{X} \rTo[l>=8em]^{\ltuple \,\ltuple \val{s_{z}} \rtuple_{z
    \in Z}\,,\,\Id_{\Cat{C}^{X}}\,\rtuple} \Cat{C}^{Z} \times
\Cat{C}^{X} \rTo[l>=4em]^{\val{t}} \Cat{C}
$$
according to the previous proposition.

\begin{prpstn}
  $\mu$-functors are closed under parameterized initial algebras and
  parameterized final coalgebras.
\end{prpstn}

By this property we mean that if $F: \Cat{C}^{Y} \rTo \Cat{C}^{X}$ is
a $\mu$-functor and $X \subseteq Y$, so that we can represent
$\Cat{C}^{Y}$ as the product $\Cat{C}^{X} \times \Cat{C}^{Y \setminus
  X}$, then we can find a collection of $\mu$-terms $\{ t_{x} \in
\mu\mathcal{T}(Y \setminus X)\}_{x \in X}$ so that $\ltuple
\val{t_{x}}\rtuple_{x \in X}: \Cat{C}^{Y \setminus X} \rTo
\Cat{C}^{X}$ is a parameterized initial algebra of $F$ and, similarly,
it is possible to find an analogous representation for a parameterized
final coalgebra of $F$.  The proposition is an immediate consequence
of the Beki\v{c} property and its dual for final coalgebras. It will
also be evident from the representation of $\mu$-functors by means of
parity functors that we describe in the next section.

We want to find concrete examples of $\mu$-bicomplete categories.  To
achieve this goal, we shall look at locally presentable categories
which, in some sense, generalize complete lattices. We briefly recall
the principal concepts that define these categories, their properties
being described in the monographs \cite{makkaipare,lpac}.

Let $\lambda$ be a regular cardinal. A poset is $\lambda$-directed if
every subset of cardinality less than $\lambda$ has an upper bound. If
$D : J \rTo \Cat{C}$ is a diagram whose index $J$ is a
$\lambda$-directed poset, then we say that $D$ is $\lambda$-directed
and that its colimit, whenever it exists, is $\lambda$-directed.  A
functor $T: \Cat{C} \rTo \Cat{D}$ is said to be $\lambda$-accessible
if it preserves $\lambda$-directed colimits.  An object $c$ of a
category $\Cat{C}$ is $\lambda$-presentable if the hom-functor
$\Cat{C}(c,-):\Cat{C} \rTo \Sets$ is $\lambda$-accessible.  Thus: a
category $\Cat{C}$ is locally $\lambda$-presentable if (1) it is
cocomplete and (2) it has a set $\mathcal{A}$ of $\lambda$-presentable
objects such that every object of $\Cat{C}$ is the $\lambda$-directed
colimit of objects from $\mathcal{A}$. We can relax condition (1) to:
(1') it has all the $\lambda$-directed colimits, in which case
conditions (1') and (2) define a $\lambda$-accessible category.
Finally: a functor is said to be \emph{accessible} if it is
$\lambda$-accessible for some regular cardinal $\lambda$.  A category
is said to be \emph{locally presentable} (\emph{accessible}) if it is
locally $\lambda$-presentable ($\lambda$-accessible) for some regular
cardinal $\lambda$.

Most of the common
categories are locally presentable: the category of sets and
functions, categories of presheaves and sheaves, varieties and
quasivarieties of algebras. Thus, in the rest of this section, we
shall prove:
\begin{thrm}
  \label{theo:lpc}
  Every locally presentable category is $\mu$-bicomplete.
\end{thrm}

In order to show that a category $\Cat{C}$ is $\mu$-bicomplete, it
suffices to find a class of functors of the form $\Cat{C}^{J} \rTo
\Cat{C}$, where $J$ ranges on finite sets, that contains the
projections and is closed under finite products, finite coproducts,
and formation of parameterized initial algebras and parameterized
final coalgebras. We recall the following facts about accessible
functors:
\begin{itemize}
\item Left and right adjoints 
  between accessible categories are accessible \cite[\S 2.23]{lpac}.
\item If $\Cat{D}$ has an initial and a final object, then a
  projection $\prj_{\Cat{C}}: \Cat{C}\times \Cat{D} \rTo \Cat{C}$ is
  both a left and a right adjoint.
\item Coproducts, diagonals and products are adjoints, since $\coprod
  \dashv \Delta \dashv \prod: \Cat{C}^{J} \rTo \Cat{C}$.
\end{itemize}
Knowing that locally presentable categories are closed under finite
products, we conclude that if $\Cat{C}$ is such a
category, then the class of accessible functors $F: \Cat{C}^{J} \rTo
\Cat{C}$ contains the projections and is closed under finite products
and finite coproducts.
It is well known that initial algebras and final coalgebras of
$\lambda$-accessible unary functors exist in locally presentable
categories \cite{MR82d:18004,barr}; moreover if $F: \Cat{C}\times
\Cat{D} \rTo \Cat{C}$ is $\lambda$-accessible, so is the unary functor
$F(-,d):\Cat{C} \rTo \Cat{C}$ for each object $d$ of $\Cat{D}$. Thus,
in order to conclude that locally presentable categories are
$\mu$-bicomplete, we need the following proposition:
\begin{prpstn}
  \label{lemma:closed}
  If $\Cat{C}$ and $\Cat{D}$ are locally presentable categories and $F
  :
  \Cat{C}\times \Cat{D} \rTo \Cat{C}$ is an accessible functor, then
  both the parameterized initial algebra $\dgmu{F} : \Cat{D} \rTo
  \Cat{C}$ and the parameterized final coalgebra $\dgnu{F} : \Cat{D}
  \rTo \Cat{C}$ are accessible.
\end{prpstn}
We are thankful to Alex Simpson for pointing out the following short
proof that relies on general properties of locally presentable
categories and accessible functors.
\begin{proof}
  We only prove that $\dgnu{F}$ is accessible, since the proof for
  $\dgmu{F}$ is dual.  Consider the category $\Cat{E}$ with objects
  $(c,d,\zeta)$, where $c \in \Cat{C}$, $d \in \Cat{D}$, and $\zeta: c
  \rTo F(c,d)$, and with morphisms $(f,g):(c,d,\zeta) \rTo
  (c',d',\zeta')$ being maps $f:c \rTo c'$ and $g:d \rTo d'$ such that
  $\zeta \comp F(f,g) = f \comp \zeta'$. Observe that there is an
  obvious forgetful functor $\Cat{E} \rTo \Cat{C}\times\Cat{D}$ as
  well as a natural transformation
  $$
  \myspecialdiagram{25}{
    []*+{\Cat{E}}="E"
    ([r(1.5)u(0.4)]*+{\Cat{C}\times\Cat{D}}="P1",
     [r(1.5)d(0.4)]*+{\Cat{C}\times\Cat{D}}="P2")
    [r(3)]*+{\Cat{C}}="C"
    "E":@/^0.5em/"P1"
    "E":@/_0.5em/"P2"
    "P2":@/_0.5em/"C"_{F}_{}="A"
    "P1":@/^0.5em/"C"^{\prj_{\Cat{C}}}_{}="B"
    "P1"[d(0.2)]="P1"
    "P2"[u(0.2)]="P2"
    "P1":@2"P2"_{\zeta}
  }
  $$
  The 2-categorical diagram above is the inserter -- cf. \cite[\S
  4]{kelly} -- of $\prj_{\Cat{C}}$ and $F$ and this implies that
  $\Cat{E}$ is accessible, since accessible categories are closed
  under lax limits \cite[\S 5.1.8]{makkaipare}. Also, it is easily
  verified that the forgetful functor $\Cat{E} \rTo
  \Cat{C}\times\Cat{D}$ creates colimits, so that $\Cat{E}$ is
  cocomplete, hence locally presentable.
  
  There is a functor $G: \Cat{D} \rTo \Cat{E}$ mapping an object $d$
  of $\Cat{D}$ to $(\dgnu{F}d,d,\zeta_{d})$ where $\zeta_{d}:\dgnu{F}d
  \rTo F(\dgnu{F}d,d)$ is a final coalgebra. Then $G$ is right adjoint
  to the accessible functor $\Cat{E} \rTo \Cat{C}\times \Cat{D} \rTo
  \Cat{D}$, hence $G$ is accessible. But $\dgnu{F}$ is simply
  $$
  \mydiagram[6em]{
    []*+{\Cat{D}}
    :[r]*+{\Cat{E}}^{G}
    :[r]*+{\Cat{C}\times\Cat{D}}
    :[r]*+{\Cat{C}}^{\prj_{\Cat{C}}}
  }
  $$
  which, as a composite of accessible functors, is accessible. 
\end{proof}

\begin{finalita}
  It is possible to directly prove proposition \ref{lemma:closed}
  along the lines of \cite{barr}. Such a proof also shows that if
  $\Cat{C}$ is a locally $\lambda$-presentable category with
  $\lambda > \omega$, then the class of
  $\lambda$-accessible functors of the form $\Cat{C}^{J} \rTo \Cat{C}$
  is closed under formation of parameterized final coalgebras.  The
  condition $\lambda > \omega$ is necessary, cf. \cite{itmo}: the
  interpretation of the $\mu$-term $\nu_{y}.(x \vee (y \land y))$ in
  the category of sets is the functor that associates to each set $X$
  the set of infinite binary trees with leaves labeled in $X$. Letting
  $X$ be the set $\nnumbers$ of natural numbers, we observe that there
  are infinitely many binary trees whose leaves are labeled by an
  infinite subset of $\nnumbers$, thus the set of this infinite binary
  tree is not the inductive limit of the sets of infinite trees whose
  leaves are labeled by a finite subset of $\nnumbers$.  Since
  $\nnumbers$ is the inductive limit of its finite subsets, we see
  that this functor is not $\omega$-accessible.  Finally, since finite
  products are $\lambda$-accessible in locally $\lambda$-presentable
  categories \cite[\S1.59]{lpac}, we can infer:
  \begin{prpstn}
    If $\lambda > \omega$, then every $\mu$-functor on a locally
    $\lambda$-presentable category is $\lambda$-accessible.
  \end{prpstn}
\end{finalita}

\begin{labripreprint} 
  We are going to give an alternative proof of proposition
  \ref{lemma:closed}, along the lines of \cite{barr}.  The proof that
  follows will show in particular that if $\Cat{C}$ is a locally
  $\lambda$-presentable category, where $\lambda > \omega$ is a
  regular cardinal, then the class of $\lambda$-accessible functors of
  the form $\Cat{C}^{J} \rTo \Cat{C}$ is closed under formation of
  parameterized final coalgebras.  The condition $\lambda > \omega$ is
  necessary, cf. \cite{itmo}: the interpretation of the $\mu$-term
  $\nu_{y}.(x \vee (y \land y))$ in the category of sets is the
  functor that associates to each set $X$ the set of infinite binary
  trees with leaves labeled in $X$. Letting $X$ be the set $\nnumbers$
  of natural numbers, we observe that there are infinitely many binary
  trees whose leaves are labeled by an infinite subset of $\nnumbers$,
  thus the set of this infinite binary tree is not the inductive limit
  of the sets of infinite trees whose leaves are labeled by a finite
  subset of $\nnumbers$.  Since $\nnumbers$ is the inductive limit of
  its finite subsets, we see that this functor is not
  $\omega$-accessible.
  
  Since it is well known that finite products are $\lambda$-accessible
  in locally $\lambda$-presentable categories, cf. for example
  \cite[\S1.59]{lpac}, this result will also imply the following fact:
  \begin{prpstn}
    If $\lambda > \omega$, then every $\mu$-functor on a locally
    $\lambda$-presentable category is $\lambda$-accessible.
  \end{prpstn}

  To prove proposition
  \ref{lemma:closed}, we rely on standard properties and
  language of locally presentable categories, cf.  \cite{lpac}.
  We begin from initial algebras.
  \begin{prpstn}
    Let $\Cat{C},\Cat{W}$ be a locally $\lambda$-presentable categories
    and $T: \Cat{C} \times \Cat{W} \rTo \Cat{C}$ be a
    $\lambda$-accessible functor. Then, for each object $w$ of $\Cat{W}$
    an initial $T(-,w)$-algebra $(\bd{x}_{w},\struct{x}_{w})$ exists and
    the induced functor $\bd{x} : \Cat{W} \rTo \Cat{C}$ is again
    $\lambda$-accessible.
  \end{prpstn}
  \begin{proof}
    It is well known that such an initial algebra exists, see for
    example \cite{MR82d:18004}.  Let $I$ be a $\lambda$-directed poset,
    $w: I \rTo \Cat{W}$ a functor with colimiting cocone
    $$
    (\inj_{w_{i}}:w_{i} \rTo w)_{i \in I}\,.
    $$
    We shall show first that $\colim \bd{x}_{w_{i}}$ has a
    $T(-,w)$-algebra structure which is moreover initial.  Recall that
    the arrow
    $$
    \lcotuple T(\inj_{\bd{x}_{w_{i}}},\inj_{w_{i}})\rcotuple : \colim
    T(\bd{x}_{w_{i}},w_{i}) \rTo T(\colim \bd{x}_{w_{i}},w)
    $$
    is invertible, since $T$ is $\lambda$-accessible, and thus
    construct a $T(-,w)$-algebra as follows:
    $$
    T(\colim \bd{x}_{w_{i}},w) \iso \colim T(\bd{x}_{w_{i}},w_{i})
    \rTo[l>=4em]^{\colim \struct{x}_{w_{i}}} \colim \bd{x}_{w_{i}}\,.
    $$
    Let $\alpha: T(a,w) \rTo a$ be a $T(-,w)$-algebra, and, for each
    $i \in I$, form the $T(-,w_{i})$-algebra
    $$
    T(a,\inj_{w_{i}}) \comp \alpha : T(a,w_{i}) \rTo T(a,w) \rTo a\,.
    $$
    Let $f_{i}: \bd{x}_{w_{i}} \rTo a$ be such that
    \begin{eqnarray*}
      \struct{x}_{w_{i}} \comp f_{i} 
      & = & 
      T(f_{i},w_{i}) \comp T(a,\inj_{w_{i}}) \comp \alpha\,,
    \end{eqnarray*}
    and argue that $f_{i} = \bd{x}_{w_{ij}}\comp f_{j}$ -- i.e.
    $(f_{i}:\bd{x}_{w_{i}} \rTo a)_{i \in I}$ is a cocone -- by means
    of the following diagram
    $$
    \mydiagrambot[6em]{ []( !S
      {T(\bd{x}_{w_{i}},w_{i})}{\bd{x}_{w_{i}}}
      {T(\bd{x}_{w_{j}},w_{j})}{\bd{x}_{w_{j}}} {1}{1.5}, !A
      {\struct{x}_{w_{i}}}{T(\bd{x}_{w_{ij}},w_{ij})}
      {\bd{x}_{w_{ij}}}{} "1"="Start" ) "3" ( !S
      {T(\bd{x}_{w_{j}},w_{j})}{\bd{x}_{w_{j}}} {T(a,w_{j})}{a}
      {1}{1.5}, !A {\struct{x}_{w_{j}}}{T(f_{j,w_{j}})}
      {f_{j}}{T(a,\inj_{w_{j}})\comp \alpha} "3"="T(a,w_{j})" ) "Start"
      :[ru]*+{T(a,w_{i})}^{T(\bd{x}_{w_{ij}} \comp f_{j},w_{i})}
      :"T(a,w_{j})"^{T(a,w_{ij})} }\,.
    $$
    The relation
    \begin{eqnarray*}
      \colim \struct{x}_{w_{i}} \comp \lcotuple f_{i}
      \rcotuple
      & = & \lcotuple T(\inj_{\bd{x}_{w_{i}}},\inj_{w_{i}})  \rcotuple
      \comp 
      T(\lcotuple f_{i} \rcotuple, a) \comp \alpha\,,
    \end{eqnarray*}
    is easily deduced as follows:
    \begin{eqnarray*}
      \lefteqn[3cm]{\inj_{T(\bd{x}_{w_{i}},w_{i})} \comp \colim \struct{x}_{w_{i}} \comp \lcotuple f_{i}
        \rcotuple}\\
      & = & \struct{x}_{w_{i}} \comp f_{i} \\
      & = & T(f_{i},w_{i})\comp T(a,\inj_{w_{i}}) \comp \alpha \\
      & = & 
      T(\inj_{\bd{x}_{w_{i}}} \comp \lcotuple f_{i} \rcotuple,\inj_{w_{i}}) \comp \alpha \\
      & = & 
      T(\inj_{\bd{x}_{w_{i}}},\inj_{w_{i}}) \comp 
      T(\lcotuple f_{i} \rcotuple, w) \comp \alpha \\
      & = & \inj_{T(\bd{x}_{w_{i}},w_{i})} \comp 
      \lcotuple T(\inj_{\bd{x}_{w_{i}}},\inj_{w_{i}})  \rcotuple
      \comp 
      T(\lcotuple f_{i} \rcotuple, w) \comp \alpha
    \end{eqnarray*}
    and hence we obtain
    \begin{eqnarray}
      \label{eq:goal}
      \lcotuple T(\inj_{\bd{x}_{w_{i}}},\inj_{w_{i}})  \rcotuple^{-1}
      \comp \colim \struct{x}_{w_{i}} \comp \lcotuple f_{i}
      \rcotuple
      & = & 
      T(\lcotuple f_{i} \rcotuple, w) \comp \alpha\,.
    \end{eqnarray}
    On the other hand, if a relation like $(\ref{eq:goal})$ with $g$ in
    place of $\lcotuple f_{i} \rcotuple$ holds, then we obtain
    \begin{eqnarray*}
      \struct{x}_{w_{i}}\comp \inj_{\bd{x}_{w_{i}}} \comp g
      & = & 
      T(\inj_{\bd{x}_{w_{i}}},\inj_{w_{i}}) \comp  
      T(g, w) \comp \alpha \\
      & = & 
      T(\inj_{\bd{x}_{w_{i}}} \comp g,w_{i}) \comp  
      T(a,\inj_{w_{i}}) \comp \alpha
    \end{eqnarray*}
    for each $i\in I$. Therefore $\inj_{\bd{x}_{w_{i}}} \comp g =
    f_{i}$ and $g = \lcotuple f_{i} \rcotuple$.
    
    In order to conclude the argument, observe that for the particular
    case of the given initial algebra $\struct{x}_{w}:T(\bd{x}_{w},w)
    \rTo w$, the $f_{i}$'s are the $\bd{x}_{\inj_{w_{i}}}$'s, since
    $\struct{x}_{w_{i}} \comp \bd{x}_{\inj_{w_{i}}} =
    T(\bd{x}_{\inj_{w_{i}}},w_{i}) \comp T(\bd{x}_{w},\inj_{w_{i}})
    \comp \struct{x}_{w}$. It follows that $\lcotuple
    \bd{x}_{\inj_{w_{i}}} \rcotuple$ is a morphism of
    $T(-,w)$-algebras between two initial ones, and henceforth it is
    invertible.
  \end{proof}
  
  We analyze next parameterized final coalgebras. 
  \begin{prpstn}
    Let $\lambda > \omega$ be a regular cardinal and $T: \Cat{C} \rTo
    \Cat{C}$ be a $\lambda$-accessible functor of a locally
    $\lambda$-presentable category.  Call $\mathcal{A}_{T}$ the class of
    $T$-coalgebras $(a,\alpha)$ such that $a$ is $\lambda$-presentable
    as an object of $\Cat{C}$. Then, up to isomorphism,
    $\mathcal{A}_{T}$ is a set and moreover it is a strong generator for
    the category of $T$-coalgebras $\Cat{C}_{T}$.
  \end{prpstn}
  We observe that the above proposition does not imply that
  $\Cat{C}_{T}$ is locally $\lambda$-presentable since a coalgebra
  $(a,\alpha) \in \mathcal{A}_{T}$ need not be $\lambda$-presentable in
  the category $\Cat{C}_{T}$.
  
  \begin{proof}
    It is easily shown that $\mathcal{A}_{T}$ is, up to isomorphism, a
    set, so that we will show that each coalgebra $(c,\gamma)$ is the
    colimit of the canonical diagram $\dom: \mathcal{A}_{T}/(c,\gamma)
    \rTo \Cat{C}_{T}$.
    
    To this end, fix the coalgebra $\gamma : c \rTo Tc$, let
    $\mathcal{A}$ be the class of $\lambda$-presentable objects in
    $\Cat{C}$ and call $U_{c}$ the lifting of the forgetful functor $U :
    \Cat{C}_{T} \rTo \Cat{C}$ to the comma categories:
    $$
    \mydiagrambot[7em]{ []( !S
      {\mathcal{A}_{T}/(c,\gamma)}{\Cat{C}_{T}} {\mathcal{A}/c}{\Cat{C}}
      {1}{1}, !A {\dom}{U_{c}} {U}{\dom} ) [] }\,.
    $$
    We shall show that the functor $U_{c}$ is cofinal, from which it
    follows
    $$
    \renewcommand{\arraystretch}{1.3}
    \begin{array}[b]{rcl@{\hspace{7mm}}l}
      c & = & \colim \dom \\
      & = & \colim (\dom \circ U_{c})
      & \textrm{since $U_{c}$ is cofinal} \\
      & = & \colim (U \circ \dom)\,,
    \end{array}
    \renewcommand{\arraystretch}{1}
    $$
    and
    \begin{eqnarray*}
      (c,\gamma) & = & \colim \dom\,,
    \end{eqnarray*}
    since the functor $U : \Cat{C}_{T} \rTo \Cat{C}$ creates colimits.
    Recall that an object of $\mathcal{A}_{T}/(c,\gamma)$ is a triple
    $(a,\alpha,f)$ such that $(a,\alpha)$ is a coalgebra in
    $\mathcal{A}_{T}$ and $f: (a,\alpha) \rTo (c,\gamma)$ is a
    coalgebra morphism; an arrow $h:(a,\alpha,f) \rTo (b,\beta,g)$ of
    $\mathcal{A}_{T}/(c,\gamma)$ is a coalgebra morphism $h:(a,\alpha)
    \rTo (b,\beta)$ such that $h \comp g = f$.  The explicit
    description of the functor $U_{c}$ is as follows: the triple
    $(a,\alpha,f)$ is sent to the pair $(a,f)$ and the arrow
    $h:(a,\alpha,f) \rTo (b,\beta,g)$ is sent to $h:(a,f) \rTo (b,g)$.
    
    We begin by showing that for a given object $(a,f)$ of
    $\mathcal{A}/c$, we can find an object $(b,\beta,g)$ of
    $\mathcal{A}_{T}/(c,\gamma)$ and an arrow $h: (a,f) \rTo (b,g)$ in
    $\mathcal{A}/c$.
    
    Recall that $c = \colim b_{j}$ where $b : J \rTo \Cat{C}$ is a
    $\lambda$-directed diagram taking image in $\mathcal{A}$.  Let
    $(b_{0},g_{0}) = (a,f)$.  Suppose that we have constructed an
    object $(b_{n},g_{n})$ of $\mathcal{A}/c$, then we can construct a
    new object $(g_{n +1},b_{n+ 1})$ of $\mathcal{A}/c$ and an arrow
    $\beta_{n} : b_{n} \rTo Tb_{n +1}$ such that the diagram
    $$
    \mydiagram[6em]{ [] ( !S {b_{n}}{c} {Tb_{n + 1}}{Tc} {1}{1}, !A
      {g_{n}}{\beta_{n}} {\gamma}{Tg_{n + 1}} ) }
    $$
    is commutative. This is possible, since $b_{n}$ is
    $\lambda$-presentable, $T$ is $\lambda$-accessible, so that
    $(T\inj_{j}:Tb_{j} \rTo Tc)_{j \in J}$ is a colimiting cocone.
  
    Consider now the coalgebra
    $$
    \coprod_{n \geq 0} b_{n} \rTo[l>=4em]^{\coprod_{n \geq 0}
      \beta_{n}} \coprod_{n \geq 0} Tb_{n + 1} \rTo[l>=4em]^{\lcotuple
      T\inj_{b_{n + 1}}\rcotuple} T(\coprod_{n \geq 0} b_{n})\,.
    $$
    Its carrier $\coprod_{n \geq 0} b_{n}$ is a $\lambda$-small
    colimit of $\lambda$-presentable objects, hence it is again
    $\lambda$-presentable.  We are using here the fact that $\lambda >
    \omega$, so that a countable colimit of $\lambda$-presentable
    objects is again $\lambda$-presentable, cf. \cite[\S 1.16]{lpac}.
    
    Moreover $\lcotuple g_{n} \rcotuple$ is a morphism of coalgebras to
    $(c,\gamma)$, by virtue of the following calculations:
    \begin{eqnarray*}
      \lefteqn[3cm]{\inj_{b_{n}} \comp 
        (\coprod_{n \geq 0} \beta_{n}  
        \comp \lcotuple T\inj_{b_{n+ 1}} \rcotuple)
        \comp T\lcotuple g_{n} \rcotuple} \\[-0.8em]
      & = & \beta_{n} \comp \inj_{Tb_{n + 1}} \comp \lcotuple T\inj_{b_{n
          + 1}} \rcotuple \comp T\lcotuple g_{n} \rcotuple \\
      & = & \beta_{n} \comp  T\inj_{b_{n
          + 1}} \comp T\lcotuple g_{n} \rcotuple \\
      & = & \beta_{n} \comp  Tg_{n + 1} \\ 
      & = & g_{n} \comp \gamma \\
      & = & \inj_{b_{n}} \comp \lcotuple g_{n} \rcotuple \comp \gamma\,.
    \end{eqnarray*}
    Summarizing, let $b = \coprod_{n \geq 0} b_{n}$, $\beta =
    \coprod_{n \geq 0} \beta_{n} \comp \lcotuple T\inj_{b_{n+ 1}}
    \rcotuple$, $g = \{g_{n} \}$, $h = \inj_{b_{0}}$, then
    $(b,\beta,g)$ is the desired object of
    $\mathcal{A}_{T}/(c,\gamma)$ and $h: (a,f) \rTo (b,g)$ is the
    desired arrow of $\mathcal{A}/c$, since $\inj_{b_{0}} \comp
    \lcotuple g_{n} \rcotuple = g_{0} = f$.
  
    We show now that given two arrows $h_{i} : (a,f) \rTo
    U_{c}(b_{i},\beta_{i},g_{i})$, $i = 1,2$, we can find an object
    $(d,\delta,k)$ and two arrows $l_{i}:(b_{i},\beta_{i},g_{i}) \rTo
    (d,\delta,k)$ such that $h_{1} \comp l_{1} = h_{2} \comp l_{2}$.
    We are given the diagram
    $$
    \mydiagram[6em]{ [] ( []*+{a}="1"
      ([dl]*+{b_{1}}="2",[d]*+{c}="4",[dr]*+{b_{2}}="3") !a
      {_{h_{1}}}{^{h_{2}}} {^{g_{1}}}{_{g_{2}}}, "2"="b1", "3"="b2" )
      "b1" ( !S {b_{1}}{Tb_{1}} {c}{Tc} {1}{1}, !A {\beta_{1}}{}
      {Tg_{1}}{\gamma} ) "b2" ( !S {b_{2}}{Tb_{2}} {c}{Tc} {1}{-1}, !a
      {^{\beta_{2}}}{} {_{Tg_{2}}}{} ) }
    $$
    where $f = h_{1} \comp g_{1} = h_{2} \comp g_{2}$.
    Form the pushouts from  $a$ 
    to obtain the diagram:
    $$
    \mydiagrambot[5.5em]{ []( !S {b_{1}}{Tb_{1}} {b_{1} \push{a}
        b_{2}} {Tb_{1} \push{a} Tb_{2}} {2}{1}, !a
      {^{\beta_{1}}}{^/-2mm/{\inj_{b_{1}}}}
      {^/-3mm/{\inj_{Tb_{1}}}}{^{\beta_{1} \push{a} \beta_{2}}}, !s
      {b_{1}}{Tb_{1}} {c}{Tc} {2}{0}{2}{1}, !a {}{_/-5mm/{g_{1}}}
      {}{^/-10mm/{\gamma}} "3"="c" "4"="Tc" ) [rrr] ( !S
      {b_{2}}{Tb_{2}} {b_{1} \push{a} b_{2}} {Tb_{1} \push{a} Tb_{2}}
      {2}{-2}, !a {^{\beta_{2}}}{_{\inj_{b_{2}}}}
      {_/-5mm/{\inj_{Tb_{2}}}}{} "3"="b_{1} \push{a} b_{2}"
      "4"="Tb_{1} \push{a} Tb_{2}" , !s {b_{2}}{Tb_{2}} {c} {Tc}
      {2}{0}{-1}{1}, !A {}{g_{2}} {}{} ) "Tb_{1} \push{a} Tb_{2}" (
      :[d]*+{T(b_{1}\push{a} b_{2})}_/3mm/{\lcotuple
        T\inj_{b_{1}},T\inj_{b_{2}} \rcotuple} :"Tc"_/2mm/{T\lcotuple
        g_{1},g_{2} \rcotuple}, :"Tc"|/-2.5mm/{\lcotuple Tg_{1},Tg_{2}
        \rcotuple} ) "b_{1} \push{a} b_{2}":"c"|/-2.5mm/{\lcotuple
        g_{1},g_{2} \rcotuple} }\,.
    $$
    We let $d = b_{1}\push{a} b_{2}$, $\delta = \beta_{1} \push{a}
    \beta_{2} \comp \lcotuple T\inj_{b_{1}},T\inj_{b_{2}} \rcotuple$,
    $k =\lcotuple g_{1},g_{2} \rcotuple$, then $(d,\delta,k)$ is an
    object of $\mathcal{A}_{T}/(c,\gamma)$:
    \begin{eqnarray*}
      \delta \comp Tk & = & 
      \beta_{1} \push{a} \beta_{2} \comp 
      \lcotuple T\inj_{b_{1}},T\inj_{b_{2}} \rcotuple \comp
      T \lcotuple g_{1},g_{2} \rcotuple \\ 
      & = & 
      \beta_{1} \push{a} \beta_{2} \comp 
      \lcotuple Tg_{1},Tg_{2} \rcotuple \\
      & = & \lcotuple g_{1},g_{2} \rcotuple \comp \gamma \\
      & = & k \comp \gamma \,.
    \end{eqnarray*}
    On the other hand, for $i = 1,2$, we let $l_{i} = \inj_{b_{i}}$.
    Then $l_{i}:(b_{i},\beta_{i},g_{i}) \rTo (d,\delta,k)$ is a
    morphism in $\mathcal{A}_{T}/(c,\gamma)$:
    \begin{eqnarray*}
      \inj_{b_{i}} \comp \delta & = & 
      \inj_{b_{i}} \comp (\beta_{1} \push{a} \beta_{2} \comp 
      \lcotuple T\inj_{b_{1}},T\inj_{b_{2}} \rcotuple) \\
      & = & \beta_{i} \comp \inj_{Tb_{i}}\comp 
      \lcotuple T\inj_{b_{1}},T\inj_{b_{2}} \rcotuple \\
      & = & \beta_{i} \comp T\inj_{b_{i}}\,, \\
      \inj_{i} \comp k & = & \inj_{i} \comp \lcotuple g_{1},g_{2} \rcotuple \\
      & = & g_{i}\,.
    \end{eqnarray*}
    Finally $h_{1} \comp l_{1} = h_{2} \comp l_{2}: (a,f) \rTo (d,k)$ is
    a commutative diagram in $\mathcal{A}/c$.
  \end{proof}

  \begin{prpstn}
    Let $\Cat{C},\Cat{W}$ be locally $\lambda$-presentable categories
    and $T: \Cat{C} \times \Cat{W} \rTo \Cat{C}$ be a
    $\lambda$-accessible functor, $\lambda$ being a regular cardinal
    strictly greater than $\omega$.  Then for each object $w$ of
    $\Cat{W}$ a final $T(-,w)$-coalgebra $(\bd{x}_{w},\struct{x}_{w})$
    exists and the induced functor $\bd{x}:\Cat{W} \rTo \Cat{C}$ is
    again $\lambda$-accessible.
  \end{prpstn}
  \begin{proof}
    The existence of the final coalgebras
    $(\bd{x}_{w},\struct{x}_{w})$ has been proved for example in
    \cite{barr}.  Let $J$ be a $\lambda$-directed poset and $w : J
    \rTo \Cat{W}$ be functor with limiting cocone
    $$
    (\inj_{w_{j}}:w_{j}\rTo w)_{j \in J}\,.
    $$
    The arrows $\struct{x}_{w_{j}}:\bd{x}_{w_{j}} \rTo^{}
    T(\bd{x}_{w_{j}},w_{j})$ are natural in $j$, so that we can take
    their colimit to construct the $T(-, w)$-coalgebra
    $$
    \colim \bd{x}_{w_{j}} \rTo[l>=4em]^{\colim\struct{x}_{w_{j}}}
    \colim T(\bd{x}_{w_{j}} , w_{j}) \rTo[l>=4em]^{\lcotuple
      T(\inj_{\bd{x}_{w_{j}}},\inj_{w_{j}}) \rcotuple} T(\colim
    \bd{x}_{w_{j}} ,w)\,.
    $$
    We claim that this is a final $T(-, w)$-coalgebra. To prove the
    claim it is enough to show that if a the carrier of a $T(-,
    w)$-coalgebra $(a,\alpha)$ is $\lambda$-presentable in $\Cat{C}$,
    then there exists a unique coalgebra morphism into the coalgebra
    defined above.  Indeed, we have shown that $\mathcal{A}_{T(-,w)}$
    is a strong generator for the category of $T(-,w)$-coalgebras, so
    that the claim follows by considering that every object of
    $\Cat{C}_{T(-,w)}$ is a colimit of objects from
    $\mathcal{A}_{T(-,w)}$.
  
    Consider therefore a coalgebra $(a,\alpha)$ in
    $\mathcal{A}_{T(-,w)}$. Then we can factor $\alpha$ as $\alpha_{j}
    \comp T(a,\inj_{w_{j}})$ for some $j \in J$, since
    $(T(a,\inj_{w_{j}}):T(a,w_{j})\rTo T(a,w))_{j \in J}$ is a
    colimiting cone, the functor $T(a, -)$ being $\lambda$-accessible.
    Let $f_{j}$ be such that $f_{j} \comp \struct{x}_{w_{j}} =
    \alpha_{j} \comp T(f_{j},w_{j})$, then $f_{j} \comp
    \inj_{\bd{x}_{w_{j}}}$ is a morphism of coalgebras, as shown in
    the next diagram:
    $$
    \mydiagrambot[7em]{ [] ( !S {a}{T(a,w_{j})}
      {\bd{x}_{w_{j}}}{T(\bd{x}_{w_{j}},w_{j})} {1}{1.4}, !a
      {^{\alpha_{j}}}{^{f_{j}}}
      {^/-1.5mm/{T(f_{j},w_{j})}}{^{\struct{x}_{w_{j}}}} "1"="a"
      "2"="T(a,w_{j})" "3"="\bd{x}_{w_{j}}" ) "T(a,w_{j})" ( !S
      {T(a,w_{j})}{T(a,w)}
      {T(\bd{x}_{w_{j}},w_{j})}{T(\bd{x}_{w_{j}},w)} {1}{1.4}, !A
      {T(a,\inj_{w_{j}})}{}
      {T(f_{j},w)}{T(\bd{x}_{w_{j}},\inj_{w_{j}})} "2"="T(a,w)"
      "4"="T(\bd{x}_{w_{j}},w)" ) "\bd{x}_{w_{j}}"="1" (
      !S{\bd{x}_{w_{j}}}{T(\bd{x}_{w_{j}},w_{j})} {\colim
        \bd{x}_{w_{j}}}{\colim T(\bd{x}_{w_{j}},w_{j})} {1}{1.4} !a
      {}{^{\inj_{\bd{x}_{w_{j}}}}}
      {^/-2mm/{\inj_{T(\bd{x}_{w_{j}},w_{j})}}}{|{\colim
          \struct{x}_{w_{j}}}} ) "2" (
      !S{T(\bd{x}_{w_{j}},w_{j})}{T(\bd{x}_{w_{j}},w)} {\colim
        T(\bd{x}_{w_{j}},w_{j})} {T(\colim \bd{x}_{w_{j}},w)} {1}{1.4}
      !a{}{} {^/-0.9em/{T(\inj_{\bd{x}_{w_{j}}},w)}} {|{\lcotuple
          T(\inj_{\bd{x}_{w_{j}}},\inj_{w_{j}}) \rcotuple}} )
      "a":@/_3em/"T(a,w)"_{\alpha} }
    $$
    
    On the other hand, suppose that $\alpha \comp T(f,w) = f \comp
    \colim \struct{x}_{w_{j}} \comp \lcotuple
    T(\inj_{\bd{x}_{w_{j}}},\inj_{w_{j}}) \rcotuple$ and choose
    liftings $\alpha_{i}$ and $f_{j}$ of $\alpha$ and $f$
    respectively, as usual, since $a$ is $\lambda$-presentable.
    Moreover, since $J$ is directed, we can assume that $i = j$.
  
    Recall that
    $$
    (T(\inj_{\bd{x}_{w_{j}}},\inj_{w_{j}}): T(\bd{x}_{w_{j}},w_{j})
    \rTo[l>=4em]^{} T(\colim \bd{x}_{w_{j}}, w))_{j \in J}
    $$
    is a colimiting cocone, hence $f_{j} \comp \struct{x}_{w_{j}}$
    and $\alpha_{j} \comp T(f_{j},w_{j})$ are two different liftings of
    the same arrow to this colimit:
    \begin{eqnarray*}
      \alpha \comp T(f,w) 
      & = & \alpha_{j} \comp T(a, \inj_{w_{j}}) \comp
      T(f_{j},w) \comp T(\inj_{\bd{x}_{w_{j}}},w)  \\
      & = &
      \alpha_{j} \comp T(f_{j},w_{j}) \comp
      T(\inj_{\bd{x}_{w_{j}}},\inj_{w_{j}}) \\[2mm]
      \alpha \comp T(f,w) & = & f \comp \colim
      \struct{x}_{w_{j}} \comp \lcotuple T(\inj_{\bd{x}_{w_{j}}},\inj_{w_{j}})\rcotuple
      \\
      & = & f_{j}\comp \inj_{\bd{x}_{w_{j}}}\comp \colim
      \struct{x}_{w_{j}} \comp \lcotuple T(\inj_{\bd{x}_{w_{j}}},\inj_{w_{j}}) \rcotuple
      \\
      & = & f_{j}\comp
      \struct{x}_{w_{j}} \comp  T(\inj_{\bd{x}_{w_{j}}},\inj_{w_{j}})\,.
    \end{eqnarray*}
  
    Therefore there exists a $k \in J$ such that $j \leq k$ and
    moreover such that
    \begin{eqnarray*}
      f_{j} \comp \struct{x}_{w_{j}} \comp T(\bd{x}_{w_{jk}},w_{jk})
      & = & \alpha_{j} \comp T(f_{j},w_{j})\comp T(\bd{x}_{w_{jk}},w_{jk})\,.
    \end{eqnarray*}
    By naturality of $\struct{x}$, we have
    \begin{eqnarray*}
      \struct{x}_{w_{j}} \comp T(\bd{x}_{w_{jk}},w_{jk})
      & = & \bd{x}_{w_{jk}} \comp \struct{x}_{w_{k}}\,,
    \end{eqnarray*}
    and therefore we obtain the relation
    \begin{eqnarray*}
      f_{j} \comp \bd{x}_{w_{jk}} \comp \struct{x}_{w_{k}}
      & = & \alpha_{j} \comp T(a,w_{jk}) \comp T(f_{j}\comp \bd{x}_{w_{jk}},w_{k})\,.
    \end{eqnarray*}
    We conclude that $f_{j} \comp \bd{x}_{w_{jk}}$ is uniquely
    determined as the unique $T(-,w_{k})$-coalgebra morphism from
    $\alpha_{j} \comp T(a,w_{jk})$ to the final one, and consequently
    also $f$ is uniquely determined, because of
    \begin{eqnarray*}
      f & = & f_{j} \comp \inj_{\bd{x}_{w_{j}}} \\
      & = & f_{j} \comp \bd{x}_{w_{jk}}\comp \inj_{\bd{x}_{w_{k}}}\,.
    \end{eqnarray*}
    In order to conclude the argument it is enough to show $\lcotuple
    \bd{x}_{\inj_{w_{j}}} \rcotuple$ is a morphism of $T(-,w)$-algebras
    between two initial ones:
    \begin{eqnarray*}
      \lefteqn[2cm]{\inj_{\bd{x}_{w_{i}}} \comp\lcotuple
        \bd{x}_{\inj_{w_{j}}} \rcotuple \comp \struct{x}_{w}} \\
      & = & \bd{x}_{\inj_{w_{j}}} \comp \struct{x}_{w} \\
      & = & \struct{x}_{w_{j}} \comp  
      T(\bd{x}_{\inj_{w_{j}}},\inj_{w_{j}})  \\
      & = & \struct{x}_{w_{j}} \comp  
      T(\inj_{\bd{x}_{w_{i}}} \comp \lcotuple
      \bd{x}_{\inj_{w_{j}}} \rcotuple,\inj_{w_{j}})  \\
      & = &  \struct{x}_{w_{j}} \comp T(\inj_{\bd{x}_{w_{i}}},\inj_{w_{i}}) 
      \comp T(\lcotuple
      \bd{x}_{\inj_{w_{j}}} \rcotuple,w) \\
      & = &  \struct{x}_{w_{j}} \comp \inj_{T(\bd{x}_{w_{i}},w_{i})} 
      \comp \lcotuple T(\inj_{\bd{x}_{w_{j}}},\inj_{w_{j}}) \rcotuple 
      \comp T(\lcotuple
      \bd{x}_{\inj_{w_{j}}} \rcotuple,w) \\
      & = &  \inj_{\bd{x}_{w_{i}}} \comp
      \colim \struct{x}_{w_{j}}  
      \comp  \lcotuple T(\inj_{\bd{x}_{w_{j}}},\inj_{w_{j}}) \rcotuple \comp T(\lcotuple
      \bd{x}_{\inj_{w_{j}}} \rcotuple,w) \,,
    \end{eqnarray*}
    so that $\lcotuple \bd{x}_{\inj_{w_{j}}} \rcotuple$ is invertible.
  \end{proof}
\end{labripreprint}

%% file: parityfunctors.tex
We have argued that $\mu$-functors are closed under parameterized
initial algebras: by the Beki\v{c} property, it becomes possible to
construct initial solutions of systems of functorial equations by
means of $\mu$-terms. The arising algebraic expressions representing
the solution of a system are however large in the dimension of the
system and are not unique.  Moreover, a large algebraic expression
could be useless for understanding its denotation in concrete
categories.  For this reason we would like to have some kind of
``smooth'' terms for the theory of $\mu$-bicomplete categories. These
terms should have a compact representation and possibly they should be
suggestive of their semantics.  To achieve this goal, the central
notion is that of parity game, cf.  for example \cite{zie,AN01}. We
define it here in a slightly generalized way.
\begin{dfntn}
  A \emph{parity game} is a tuple $G = \langle S,\hg,\kappa
  ,\epsilon\rangle$, where
  \begin{itemize}
  \item $S =\langle \dom,\cod : M \rTo P \rangle$ is a finite graph of
    positions and moves.
  \item $\hg: P \rTo \agset{n}$ is a function such that, if $\hg(p)
    = \omega$, then $\{ \,m\,| \,\dom(m) = p \,\} = \emptyset$.  
  \item $\kappa : \set{n} \rTo \{ \mu, \nu \}$.
  \item $\epsilon: \{\, p \in P \,| \,\hg(p) \neq \omega \,\} \rTo \{
    \sigma,\pi \}$.
  \end{itemize}
\end{dfntn}

We fix some terminology and notation. If $\hg : P \rTo \agset{n}$,
then we shall say that $n$ is the \emph{height} of $G$ and write
$\height{G} = n$.  For each $p \in P$, we let $M_{p}$ be the set
$\dom^{-1}(p)$. We let $P_{i} = \hg^{-1}(i)$, $P_{<i} = \bigcup_{j <
  i} P_{j}$, $P_{\leq i} = \bigcup_{j \leq i} P_{j}$ for $i \in
\agset{n}$. We shall also use the notation $P_{\geq i}$ for the set
$\bigcup_{j \geq i} P_{j}$. Unless specified we will assume that the
underlying structure of a given parity game $G$ is the tuple $\langle
S,\hg,\kappa ,\epsilon\rangle$, $S$ being the graph $\langle
P,M,\dom,\cod\rangle$. A pointed parity game is a pair $\langle G, p
\rangle$ where $G$ is a parity game and $p \in P$.

\vspace{0.5em}

We interpret the above data as a two person game $G(E)$, parameterized
in a choice of sets $E = \{ E_{x} \}_{x \in P_{\omega}}$.  The graph
$S$ is a board with a set of positions $P$ and a set of allowed moves
$M$. A move $m \in M$ is from position $\dom(m)$ to position
$\cod(m)$; observe that we allow different moves relating the same
pair of positions; also, the two players need not to alternate.  From
a position $p$ the set of moves $M_{p}$ is available and player
$\epsilon(p)$ among players $\sigma$ and $\pi$ must choose how to
move.  The normal play condition holds: if a player cannot move, then
he loses.  On an infinite play $\gamma = \gamma_{0} \rightarrow
\gamma_{1}\rightarrow \ldots \gamma_{n} \rightarrow \ldots $ we will
be able to find regions among $P_{1},\ldots ,P_{n}$ which are visited
infinitely often, and among them we will be able to pick a region
$P_{i}$ with $i$ maximal.  Then, this infinite path is a win for
player $\sigma$ if and only if $i$ is colored by $\nu$. More formally,
if we let
\begin{eqnarray}
  \label{eq:in}
  \In \gamma & = & \{\, i \in \set{n} \,|\,\card \{\, l\,|\,\hg(
  \gamma_{l} ) = i \,\} = \omega \,\}\,, 
\end{eqnarray}
then $\gamma$ is a  win for player $\sigma$ if and only if
\begin{eqnarray*}
\kappa(\, \max \In \gamma\,) & = & \nu\,.
\end{eqnarray*}
If a play ends in a position $x \in P_{\omega}$, then player $\sigma$
must choose an element $e \in E_{x}$, and then he wins. If $E_{x} =
\emptyset$, then he loses.

We remark that if $P_{\omega} = \emptyset$, then the above data and
game theoretic interpretation coincide with the usual one \cite[\S
4]{AN01}.  From a game theoretic point of view, we are allowed to
normalize the functions $\hg$ and $\kappa$, so that we can always
suppose that $\kappa(i) = \nu$ if and only if $i$ is odd. In this case
an infinite play $\gamma$ is a win for player $\sigma$ if and only if
the region visited infinitely often in $\gamma$ of maximal height is
odd, whence the name of ``parity game''.

In the theory of automata recognizing infinite objects the way of
specifying a set of infinite paths in a graph is called an acceptance
condition.  The acceptance condition by which we specify the set of
infinite winning plays for player $\sigma$ was introduced by Mostowski
in \cite{MR827531} and is also known as a Rabin chain condition. Under
the hypothesis that $\kappa(i) = \nu$ if and only if $i$ is odd and
that $\height{G} = 2n$, we can let $F_{k} = P_{\geq 2k-1}$ and $E_{k}
= P_{\geq 2k}$ for $k =1,\ldots ,n$. These sets form a decreasing
chain and moreover the pairs $(F_{k},E_{k})_{k = 1,\ldots ,n}$ are
Rabin pairs for the set of infinite winning plays for player $\sigma$,
meaning that an infinite path $\gamma$ is a win for player $\sigma$ if
and only if there exists $k \in \{1,\ldots ,n \}$ such that $\In
\gamma \cap F_{k} \neq \emptyset$ and $\In \gamma \cap E_{k} =
\emptyset$. There are other ways of characterizing this acceptance
condition by means of Muller tables \cite{MR95a:90238}, but we won't
investigate this subject further. On the other hand we recall that
parity games are essential tools for model checking. For example, it
is shown in \cite{MR1826118} that the problem of deciding whether
player $\sigma$ has a winning strategy from a given position of a
parity game is equivalent -- under linear reduction -- to the problem
of deciding whether a $\mu$-calculus formula holds in a given model.

\begin{figure}[h]
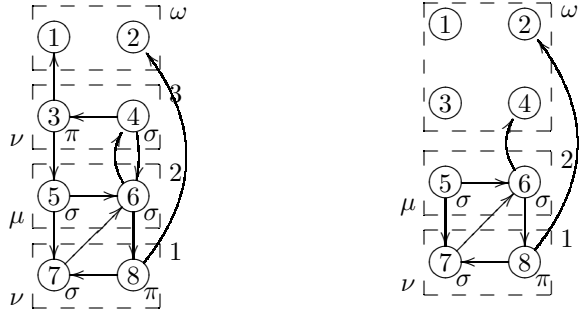

  \centering
  $$
  \mygame[3em]{
    []!E{1}([r]!E{2}="A") [d]!O{3}([r]!P{4}) [d]!P{5}([r]!P{6})
    [d]!P{7}([r]!O{8}) "3"(:"1",:"5") "5"(:"7",:"6") "7"(:"6")
    "4"(:"3",:@/^0.2em/"6") "6"(:"8",:@/^0.7em/"4")
    "8"(:"7",:@/_2em/"2") "1"("2"!L{}{\omega}) "3"("4"!L{\nu}{3})
    "5"("6"!L{\mu}{2}) "7"("8"!L{\nu}{1}) }
  \hspace{4em}
  \mygame[3em]{ []!E{1}([r]!E{2}="A") [d]!E{3}([r]!E{4})
    [d]!P{5}([r]!P{6}) [d]!P{7}([r]!O{8}) 
    "5"(:"7",:"6") "7"(:"6") 
    "6"(:"8",:@/^0.7em/"4") "8"(:"7",:@/_2em/"2") "1"("4"!L{}{\omega})
    "5"("6"!L{\mu}{2}) "7"("8"!L{\nu}{1}) }
  $$
  \vspace{-11mm}
  \caption{On the left a parity game, on the right its 
    predecessor game.}
  \label{fig:P(G)}
\end{figure}

The goal of adding positions at infinite height is to make it possible
to analyze parity games inductively. The main tool for this is the
predecessor game of a parity game, whose construction we illustrate in
figure \ref{fig:P(G)}. The game on the right is obtained from the one
on the left by erasing all the moves from the region of maximal finite
height.

\begin{dfntn}
  If $G = \langle S,\hg,\kappa ,\epsilon\rangle$ is a parity game of
  height $n > 0$, then its \emph{predecessor game} $P(G)$, of height
  $n-1$, is obtained from $G$ by erasing all the moves from $P_{n}$.
  More precisely, $P(G) = \langle S',\hg',\kappa',\epsilon' \rangle$,
  where:
  \begin{itemize}
  \item $S' = \langle \dom,\cod: \dom^{-1}(P_{< n})
    \rTo P\rangle$.
  \item $\hg'(p) = \hg(p)$ if $\hg(p) < n$, otherwise $\hg'(p) =
    \omega$.
  \item For $i \in \set{n-1}$, we let $\kappa'(i) = \kappa(i)$.
  \item If $\hg'(p) < n$, then we let $\epsilon'(p) = \epsilon(p)$.
  \end{itemize}
\end{dfntn}

In the following we shall endow the data defining a parity game with
an algebraic meaning. We let $\Cat{C}$ be a fixed category with finite
products and finite coproducts.  If $G$ is a parity game, then for
each $p \in P_{< \omega}$ we let
$$
\begin{array}{cclcl}
  \prj(\cod,p)
  & = &  \ltuple \prj_{\cod(m)}
  \rtuple_{ m \in M_{p} }
  &:& \Cat{C}^{P} \rTo
  \Cat{C}^{M_{p}} \\[2mm]
  \EQ_{p} & = & 
  \left \{
    \begin{array}{@{\hspace{0.5em}}ll}
      \prod \circ \prj(\cod,p)\,,&\epsilon(p) =
      \pi\\[2mm]
      \coprod \circ \prj(\cod,p) \,,&\epsilon(p) = \sigma
    \end{array}\right.
  &:& \Cat{C}^{P} \rTo
  \Cat{C} \,.
\end{array}
$$
For $k = 1,\ldots ,\height{G}$ we let
$$
\begin{array}{cclcl}
  \EQ_{k} & = & 
  \ltuple \EQ_{p}\rtuple_{p \in P_{k}}
  &: & \Cat{C}^{P} \rTo
  \Cat{C}^{P_{k}}
\end{array}
$$
and finally we let $\EQ_{G} = \EQ_{\height{G}}$.

\begin{dfntn}
  We define a partial correspondence $\val{-}$,  mapping a parity
  game $G$ to a functor $\val{G}:\Cat{C}^{P_{\omega}} \rTo
  \Cat{C}^{P_{< \omega}}$, by induction on the height, as follows.  If
  $\height{G} = 0$, then $P_{< \omega} = \emptyset$ so that there is a
  unique choice of $\val{G}$.  Suppose that $\height{G} = n > 0$ and
  that $\val{P(G)}$ is defined. Let
    \begin{eqnarray*}
      F & = &\val{P(G)}\circ \prj_{\Cat{C}^{P_{n}}\times \Cat{C}^{P_{\omega}}},
    \end{eqnarray*}
    and consider the functor
    $$
    \Cat{C}^{P_{< n }} \times \Cat{C}^{P_{n}}\times
    \Cat{C}^{P_{\omega}} \rTo[l>=4em]^{\ltuple F,\, \EQ_{G} \rtuple }
    \Cat{C}^{P_{< n}} \times \Cat{C}^{P_{n}}\,.
    $$
    If $\kappa(n) = \mu$, then we let $\val{G}$ be the
    parameterized initial algebra of the above functor, otherwise, if
    $\kappa(n ) = \nu$, we let $\val{G}$ be its parameterized final
    coalgebra. If $\val{P(G)}$ is undefined or if the required initial
    algebras or final coalgebras do not exist, then $\val{G}$ is
    undefined. 
    We say that $\Cat{C}$ is \emph{complete with respect
      to parity games} if for each parity game $G$, the functor
    $\val{G}:\Cat{C}^{P_{\omega}} \rTo \Cat{C}^{P_{< \omega}}$ is
    defined. 
\end{dfntn}

Whenever the functor $\val{G}: \Cat{C}^{P_{\omega}} \rTo
\Cat{C}^{P_{<\omega}}$ is defined, it is useful to extend it to a
functor $\sval{G}:\Cat{C}^{P_{\omega}} \rTo \Cat{C}^{P}$, in the
obvious way, by setting
$$
\sval{G} = \langle \val{G},\id_{\Cat{C}^{P_{\omega}}}\rangle
: \Cat{C}^{P_{\omega}} \rTo \Cat{C}^{P_{<\omega}}\times \Cat{C}^{P_{\omega}}\,.
$$
Observe that, according to proposition \ref{prop:bekic3}, the
functor $\prj_{\height{G}} \circ \val{G}$ is a parameterized initial
algebra (or final coalgebra) of the functor $\EQ_{G}\circ
\sval{P(G)}$. Moreover, according to the same proposition, the value
of $\val{G}$ is completely determined up to natural isomorphism by
$\prj_{\height{G}} \circ \val{G}$ and $\val{P(G)}$. Hence, in order to
prove that $\val{G}$ and $\val{H}$ are naturally isomorphic, it is
enough to prove that $\EQ_{G}$ and $\EQ_{H}$ are naturally isomorphic,
that $\kappa(\height{G}) = \kappa(\height{H})$, and that $\val{P(G)}$
is naturally isomorphic to $\val{P(H)}$.

\begin{dfntn}
  We say that a functor $F: \Cat{C}^{I} \rTo \Cat{C}^{J}$ is a
  \emph{parity functor} if it is naturally isomorphic to a functor of
  the form $\prj_{J} \circ \sval{G}$, where $G$ is a parity game such
  that $P_{\omega} = I$ and $J \subseteq P$ is a subset of positions.
\end{dfntn}
If in the previous lemma $I = \{p \}$ is a singleton, we will use the
notation $\sval{G}_{p}$ for the functor $\prj_{p} \circ \sval{G}$.

\begin{figure}[h]
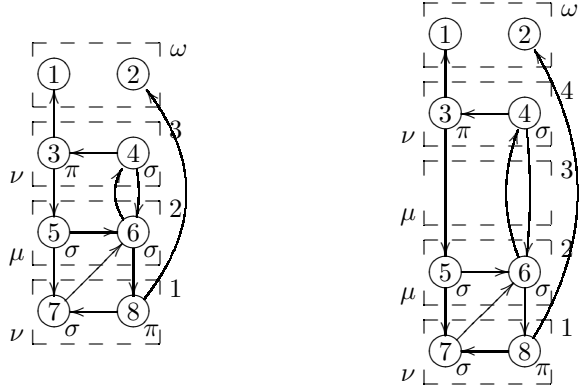

  \centering
  $$
  \mygame[3em]{
    []!E{1}([r]!E{2}="A") [d]!O{3}([r]!P{4}) [d]!P{5}([r]!P{6})
    [d]!P{7}([r]!O{8}) "3"(:"1",:"5") "5"(:"7",:"6") "7"(:"6")
    "4"(:"3",:@/^0.2em/"6") "6"(:"8",:@/^0.7em/"4")
    "8"(:"7",:@/_2em/"2") "1"("2"!L{}{\omega}) "3"("4"!L{\nu}{3})
    "5"("6"!L{\mu}{2}) "7"("8"!L{\nu}{1}) }
  \hspace{4em}
  \mygame[3em]{
    []!E{1}([r]!E{2}="A") 
    [d]!O{3}([r]!P{4}) [d(2)]!P{5}([r]!P{6})
    [d]!P{7}([r]!O{8})
    "3"[d]!I{9}([r]!I{10})
    "3"(:"1",:"5") "5"(:"7",:"6") "7"(:"6")
    "4"(:"3",:@/^0.2em/"6") "6"(:"8",:@/^0.7em/"4")
    "8"(:"7",:@/_2em/"2") "1"("2"!L{}{\omega}) "3"("4"!L{\nu}{4})
    "5"("6"!L{\mu}{2}) "7"("8"!L{\nu}{1})
    "9"("10"!L{\mu}{3})
  }
  $$
  \vspace{-11mm}
  \caption{A game $G$ on the left and 
    the game $G_{3,\mu}$ on the right.}
  \label{fig:suspension}
\end{figure}

In the following two lemmas, needed in the proof of proposition
\ref{prop:termstogames}, we exemplify how game theoretical ideas lift
to the algebra.  In \ref{def:suspension} we introduce two
constructions which respectively introduce and eliminate holes in the
height. The first construction is exemplified in figure
\ref{fig:suspension}. In \ref{def:normalized} and
\ref{lemma:normalized} we show that regions of contiguous heights can
always be assumed to be non empty and have different colors $\{\mu,\nu
\}$.  These constructions are shown to be algebraic invariants.

In the following definition, let $\hat{\imath} : \{1,\ldots ,n\} \rTo
\{1,\ldots ,n + 1 \}$ be the unique order preserving injection which
avoids $i \in \{1,\ldots ,n + 1\}$.
\vfill\eject\noindent
\begin{dfntn}
  \label{def:suspension}
  Let $G = \langle S,\hg,\kappa,\epsilon \rangle$ and suppose  that
  $\height{G} = n$. 
    \begin{itemize}
  \item For $i = 1,\ldots,n + 1$ and $\theta \in \{\mu,\nu \}$, we
    define $G_{i,\theta} = \langle
    S,\hg_{i},\kappa_{i,\theta},\epsilon \rangle$ by letting
    $\hg_{i} = \hg \comp \hat{\imath}$ and , $\kappa_{i,\theta}(j) =
    \kappa(j)$ if $j < i$, $\kappa_{i,\theta}(i) = \theta$ and
    $\kappa_{i,\theta}(j) = \kappa(j-1)$ if $j > i$.
  \item We define $G_{\bullet} = \langle
    S,\hg_{\bullet},\kappa_{\bullet},\epsilon \rangle$ as follows: we
    let $\hg_{\bullet} \comp j$ be the unique factorization of $\hg$
    such that $\hg_{\bullet}: P \setminus P_{\omega} \rOnto \{1,\ldots
    ,k \}$ is surjective and $j :\{1,\ldots ,k \} \rInto \{1 ,\ldots
    ,n \}$ is injective and order preserving; we let
    $\kappa_{\bullet} = j \comp \kappa$.
  \end{itemize}
\end{dfntn}
Observe that $\height{G_{i,\theta}} = n + 1$ and that, in the game
$G_{\bullet}$, $P_{j} \neq \emptyset$ for $j = 1,\ldots ,
\height{G_{\bullet}}$.
\begin{lmm}
  \label{lemma:suspensions}
  There exist natural isomorphisms $\val{G} \iso
  \val{G_{i,\theta}}$ and $\val{G} \iso \val{G_{\bullet}}$.
\end{lmm}
\begin{proof}
  The isomorphism $\val{G_{n + 1,\mu}} \iso
  \val{G}$
  follows by observing that $P(G_{n+1,\mu}) 
  \linebreak 
  = G$ and by letting
  $\Cat{C}$ be $\Cat{C}^{P_{\leq n}}$, $\Cat{D} = \Cat{C}^{P_{n + 1}}
  = \Cat{C}^{\emptyset} = 1$, $\Cat{E} = \Cat{C}^{P_{\omega}}$ in
  proposition \ref{prop:bekic2}: the left projection of $\val{G_{n +
      1,\mu}}: \Cat{C}^{P_{\omega}} \rTo \Cat{C}^{P_{\leq n}} \times
  1$ is computed as $\val{G}$. An analogous observation shows that
  there is an isomorphism $\val{G_{n + 1,\nu}} \iso \val{G}$.
  
  If $i \leq n$, then we can reason by induction on the height,
  observing that $P(G_{i,\theta}) = P(G)_{i,\theta}$ and
  $\EQ_{G_{i,\theta}} = \EQ_{G}$, so that $\EQ_{G_{i,\theta}} \circ
  \sval{P(G_{i,\theta})}$ and $\EQ_{G} \circ \sval{P(G)}$ are
  naturally isomorphic.
  
  On the other hand, we argue that $\val{G}$ and $\val{G_{\bullet}}$
  are naturally isomorphic as follows: $j$ can be factored by a
  sequence of the functions $\hat{\imath}$, hence $G$ can be obtained
  from $G_{\bullet}$ by a sequence of the operations
  $(-)_{i,\theta}$ and the result follows from our previous
  considerations.  
\end{proof}

\begin{dfntn}
  \label{def:normalized}
  We say that a parity game $G$ is \emph{normalized} if $\kappa(i)
  \neq \kappa(i + 1)$ for $i = 1,\ldots ,\height{G}-1$ and $P_{i} \neq
  \emptyset$ for $i = 1,\ldots ,\height{G}$.
\end{dfntn}
\begin{lmm}
  \label{lemma:normalized}
  For each parity game $G$ there exists a normalized parity game
  $N(G)$ on the same set of positions of $G$ such that $\val{G} \iso
  \val{N(G)}$.
\end{lmm}
\begin{proof}  
  Let $G = \langle S,\hg,\kappa,\epsilon \rangle$ be a game with
  $\height{G} = n + 1 > 1$. We first define a game $G_{N}$.  If
  $\kappa(n + 1) = \kappa(n)$, then we let $G_{N} = \langle
  S,\hg_{N},\kappa,\epsilon \rangle$, where $\hg_{N}(p) = n$ if
  $\hg(p) = n + 1$ and otherwise $\hg_{N}(p) = \hg(p)$. To verify that
  $\val{G}$ is isomorphic to $\val{G_{N}}$, observe that $P(G_{N}) =
  P(P(G))$ and that $\EQ_{G_{N}} = \ltuple \EQ_{P(G)},\EQ_{G} \rtuple$
  and therefore let $F = \ltuple \val{P(P(G))} \circ \prj_{P_{\geq
      n}}, \EQ_{P(G)}\rtuple$ and $G = \EQ_{G}$ in the statement of
  the Beki\v{c} property \ref{prop:bekic}.  Otherwise, if $\kappa(n +
  1) \neq \kappa(n)$, then we let $G_{N} = G$.
  
  We define then $N(G)$ by induction on the height.  If $\height{G}
  \leq 1$, then $N(G) = G$. Otherwise, in order to obtain $N(G)$, we
  first construct $N(P(G))$, and then a game $G'$ by adding to the
  region of maximal height of $G$ transitions so that $\EQ_{G'} =
  \EQ_{G}$. As the last step, we let $N(G) = G'_{N}$. By induction, it
  is shown that $\val{N(G)} = \val{G}$.  
\end{proof}

The following theorem is the main result of this section and
generalizes to categories the well known fact that a vectorial
$\mu$-calculus has no more expressive power of its scalar version
\cite[\S 2.7]{AN01}.
\begin{thrm}
  \label{prop:iff}
  A category is $\mu$-bicomplete with respect to parity games if and
  only if it is $\mu$-bicomplete.
\end{thrm}
In order to prove the theorem we translate parity functors into
collections of $\mu$-terms and vice-versa we represent $\mu$-terms
by pointed parity games.  To show that this translation is sound the
main tool is the Beki\v{c} property discussed in section
\ref{sec:bekic}.

\begin{prpstn}
  \label{prop:gamestoterms}
  For each parity game $G$ we can find a collection of $\mu$-terms $\{
  s_{p} \}_{p \in P}$, such that $s_{p} \in
  \mu\mathcal{T}(P_{\omega})$ and
  \begin{eqnarray*}
    \sval{G} & := & \ltuple \,\val{s_{p}} \,\rtuple_{p \in P}\,.
  \end{eqnarray*}
\end{prpstn}
The meaning of the symbol $:=$ is that the functorial expression on
the right determines the existence of the functorial expression on the
left. That is, natural transformations (needed as projections,
injections and as the structure part of initial algebras or final
coalgebras) can be constructed out of the natural transformations
given with the interpretations of the $\mu$-terms, so that the
functorial expression on the right together with these new natural
transformations have the universal property that determines the
left-hand side of the equation up to canonical isomorphism. Thus it
follows:
\begin{crllr}
  If $\Cat{C}$ is a $\mu$-bicomplete category, then
  $\Cat{C}$ is  complete w.r.t. parity games.
\end{crllr}
\begin{proof}[Proof of proposition \ref{prop:gamestoterms}]
  Clearly, it is enough to find a collection of $\mu$-terms indexed by
  $P_{< \omega}$ such that $\val{G} := \ltuple \,\val{s_{p}}
  \,\rtuple_{p \in P_{< \omega}}$, since then we can complete this
  collection to a collection representing $\sval{G}$, by letting
  $s_{p}$ be the $\mu$-term $p \in \mu\mathcal{T}(P_{\omega})$ if $p
  \in P_{\omega}$.

  If  $\height{G} = 0$, then the statement is true
  since $P_{ < \omega} = \emptyset$ and the empty collection of terms
  satisfies the requirements.
  
  Suppose that $\height{G} = n > 0$ and that that $\kappa(n) = \mu$.
  An analogous argument works if $\kappa(n) = \nu$.
  
  By the induction hypothesis there are $\mu$-terms $\{ s_{p} \}_{p
    \in P}$ with $s_{p} \in \mu\mathcal{T}(P_{n} \cup
  P_{\omega})$ such that $\sval{P(G)} = \ltuple \,\val{s_{p}}
  \,\rtuple_{p \in P}$.
  According to proposition \ref{prop:bekic2}, we can construct the
  functor $\val{G}$ by means of $\mu$-terms, provided we are able to
  show that the functor $\EQ_{G} \circ \sval{P(G)}$
  admits a parameterized initial algebra which is representable by
  means of $\mu$-terms. 
  We prove this by induction on the cardinality of $P_{n}$.
  If $P_{n} = \emptyset$, then there is nothing to prove.
  Otherwise, pick $p_{0} \in P$, let $P'_{n} = P_{n} \setminus \{
  p_{0} \}$ and represent the functor $\EQ_{G}$ as $\langle \EQ_{P'_{n}}, \EQ_{p_{0}} \rangle$
  where $\EQ_{P'_{n}} = \ltuple \, \EQ_{p} \, \rtuple_{p \in P'_{n}}$.
  We claim that an initial algebra of the functor $\EQ_{P'_{n}} \circ
  \sval{P(G)}$ exists and is constructible by means of $\mu$-terms.
  Indeed a parity game $G'= \langle S' ,\hg',\kappa'
  ,\epsilon'\rangle$ on the same set of positions and with the same
  height as $G$, such that $\kappa'(i) = \kappa(i)$ for $i =1,\ldots
  ,\height{G}$, $P(G') = P(G)$, $\hg'(p) = n $ if and only if $p \in
  P'_{n}$ and $\EQ_{G'} = \EQ_{P'_{n}}$, is easily constructed out of
  $G$.
  Since $\card P'_{n} < \card P_{n}$ by the induction hypothesis we
  have a desired representation of $\val{G'}$ by $\mu$-terms $t_{p}
  \in \mu\mathcal{T}(\{ p_{0} \} \cup P_{\omega})$, for $p \in P_{\leq
    n}\setminus \{p_{0}\}$. It follows that $\langle
  \val{t_{p}}\rangle_{p \in P_{n}'}$ is the desired representation of
  the initial algebra of $\EQ_{P'_{n}} \circ \sval{P(G)}$, since by
  proposition \ref{prop:bekic3} $\prj_{P'_{n}} \circ \sval{G'}$ is an
  initial algebra for $\EQ_{G'} \circ \sval{P(G')} = \EQ_{P'_{n}}
  \circ \sval{P(G)}$.

  Let
  $s : M_{p_{0}} \rTo \mu\mathcal{T}(P_{n} \cup P_{\omega})$ be
  the function defined by the relation
  \begin{eqnarray*}
    s(m) & = & s_{\cod m}
  \end{eqnarray*}
  and let $u \in \mu\mathcal{T}(P_{n} \cup
  P_{\omega})$ be the $\mu$-term defined as
  \begin{eqnarray*}
    u & = & 
    \left \{
      \begin{array}{cc}
        \Land[M_{p_{0}}] s\,, & \epsilon(p_{0}) = \pi \,\\
        \Lor[M_{p_{0}}] s\,, & \epsilon(p_{0}) = \sigma \,,
    \end{array}
    \right .
  \end{eqnarray*}
  then 
  \begin{eqnarray*}
    \EQ_{p_{0}} \circ \sval{P(G)}
    & = & \val{u} :
  \Cat{C}^{P_{n}}\times \Cat{C}^{P_{\omega}} \rTo \Cat{C}\,.
  \end{eqnarray*}

  We can now construct an initial algebra of $\EQ_{G} \circ
  \sval{P(G)}$ according to the Beki\v{c} property.  Let
  \begin{eqnarray*}
    v_{p_{0}} & = &  \mu_{p_{0}}.(\,u[
    t_{p}/p]_{p \in P'_{n}}\,)
  \end{eqnarray*}
  and for $p \in P'_{n}$ let
  \begin{eqnarray*}
    v_{p} & = & t_{p}[v_{p_{0}}/p_{0}]
  \end{eqnarray*}
  then the functor $\langle \val{v_{p}}\rangle_{p \in
    P_{n}}$  carries a canonical structure of an initial algebra for
  the functor $\EQ_{P_{n}} \circ \sval{P(G)}$.  
\end{proof}

\begin{prpstn}
  \label{prop:termstogames}
  For each $\mu$-term $s \in\mu\mathcal{T}(X)$ there exists a
  pointed parity game $\langle G,p\rangle$ such that $P_{\omega} = X$
  and 
  \begin{eqnarray*}
    \val{s} & := & \sval{G}_{p}\,.
  \end{eqnarray*}
\end{prpstn}
Again, the meaning of the symbol $:=$ is that the functorial
expression on the right can be endowed with a structure so that it has
the universal property which determines the left-hand side of the
equation up to canonical isomorphism. Thus it follows:
\begin{crllr}
  If $\Cat{C}$ is a category complete w.r.t. parity games, then
  $\Cat{C}$ is $\mu$-bicomplete.
\end{crllr}
\begin{lmm}
  Let $G$ be a parity game, we define $G^{p_{0}}$ to be the game
  obtained from $G$ by adding a new position $p_{0}$ to set of
  position at infinity $P_{\omega}$. Then there exists a natural
  isomorphism $\val{G^{p_{0}}} \iso \val{G}\circ
  \prj_{\Cat{C}^{P_{\omega}}}$.
\end{lmm}
\begin{proof}
  The observation is obvious if $\height{G} = 0$. On the other hand,
  if $\height{G} > 0$, then $P(G^{p_{0}}) = P(G)^{p_{0}}$ and
  $\val{P(G^{p_{0}})} \iso \val{P(G)} \circ \prj_{n,\omega}$, by
  induction. Moreover $\EQ_{G^{p_{0}}} = \EQ_{G} \circ
  \prj_{P}$, so that $\val{G^{p_{0}}}$ is the defined to be the
  initial algebra of the functor $\langle \val{P(G)} \circ
  \prj_{n,\omega} \circ \prj_{n,\omega,p_{0}}, \EQ_{G^{p_{0}}} \rangle
  = \langle \val{P(G)} \circ \prj_{n,\omega}, \EQ_{G} \rangle \circ
  \prj_{P}$.  In order to conclude the argument, observe that
  the parameterized initial algebra of a functor of the form $F \circ
  \prj_{\Cat{C}\times \Cat{D}}:\Cat{C}\times\Cat{D}\times\Cat{E} \rTo
  \Cat{C}$ has the form $\dgmu{F} \circ \prj_{\Cat{D}}$, where $\dgmu{F} :
  \Cat{D}\rTo \Cat{C}$ is the parameterized initial algebra of
  $F:\Cat{C}\times \Cat{D}\rTo \Cat{C}$.  
\end{proof}
\begin{proof}[Proof of proposition \ref{prop:termstogames}]
  By induction on the structure of $\mu$-terms.
  
  For the $\mu$-term $x$ in context $X$, we let $G$ be the parity game
  of height $0$ on the set of positions $X$, with distinguished
  position  $x \in X$.
  
  We analyze now the case of a term of the form $\Land[I] s$. By
  duality, we implicitly analyze the case of a term of the form
  $\Lor[I] s$.

  We first show that given a parity game $G = \langle S,\hg,\kappa,
  \epsilon \rangle$ and a subset $I \subseteq P$ of positions, it is
  possible to find a pointed parity game $\langle G', p_{0} \rangle$
  such that $\sval{G'}_{p_{0}} \iso \prod_{i \in I} \sval{G}_{i}$.  We
  define $G' = \langle S',\hg',\kappa', \epsilon' \rangle$ as follows:
  $S'$ is obtained from $S$ by adding a new position $p_{0}$ and moves
  $p_{0} \rightarrow i$ for each $i \in I$, $\hg'(p_{0}) = \height{G}
  + 1$ and $\hg'(p) = \hg(p)$ otherwise, $\kappa'(\height{G} + 1) =
  \mu$, and $\kappa'(j) = \kappa(j)$ if $j \leq \height{G}$,
  $\epsilon'(p_{0}) = \pi$ and $\epsilon'(p) = \epsilon(p)$ otherwise.
  We could also have set $\kappa'(\height{G} + 1) = \nu$,
  leading to an equivalent construction.

  Observe that $P(G') = G^{p_{0}}$, therefore
  \begin{eqnarray*}
    \val{P(G')} \circ \prj_{\Cat{C}^{P'_{\omega}}} 
    & = &
    \val{G^{p_{0}}} \circ \prj_{\Cat{C}^{\{ p_{0} \}}\times \Cat{C}^{P_{\omega}}} \\
    & \iso &
    \val{G} \circ \prj_{\Cat{C}^{P_{\omega}}} \circ \prj_{\Cat{C}^{\{ p_{0} \}}\times
      \Cat{C}^{P_{\omega}}}  \\
    & \iso & \val{G} \circ \prj_{\Cat{C}^{P_{\omega}}}\circ
    \prj_{\Cat{C}^{P_{<\omega}}\times \Cat{C}^{P_{\omega}}}\,, 
  \end{eqnarray*}
  and remark that an initial algebra for $\val{G} \circ
  \prj_{\Cat{C}^{P_{\omega}}}$ is exactly $\val{G}$.  Similarly
  \begin{eqnarray*}
    \mathcal{E}_{p_{0}}  & = &
    \prod \circ \prj(\cod,p_{0}) \\
    & \iso & (\prod_{i \in I}) \circ
    \prj_{\Cat{C}^{P_{<\omega}} \times \Cat{C}^{P_{\omega}}}\,.
  \end{eqnarray*}
  Using proposition \ref{prop:bekic2} (switch the roles of $F$ and
  $G$), compute an initial algebra of a functor of the form $\langle F
  \circ \prj_{\Cat{C}\times\Cat{E}}, G \circ
  \prj_{\Cat{C}\times\Cat{E}} \rangle: \Cat{C} \times \Cat{D} \times
  \Cat{E}\rTo \Cat{C} \times \Cat{D}$ as $\langle \dgmu{F}, G\circ
  \dgmu{F} \rangle$, $\dgmu{F}$ being the initial algebra of $F$.  In this
  formula let $F$ be $\val{G} \circ \prj_{\Cat{C}^{P_{\omega}}}$ and
  $G$ be $(\prod_{i \in I})$. It follows that $\sval{G'}_{p_{0}}$, the
  right projection in this formula, is $\sval{G'}_{p_{0}} =
  \prj_{p_{0}}\circ \val{G'} \iso \prod_{i \in I} \circ \prj_{i} \circ
  \val{G} = \prod_{i \in I} \sval{G}_{i}$.
  
  We come back to the original problem of finding a representation of
  the functor $\Land[I] s$ as a parity functor. Observing that we have
  solved the case of representing $\Land[\emptyset]$ in the previous
  discussion, we can suppose without loss of generality that $I =
  \{l,r\}$. Let $\langle G^{l},p^{l}\rangle$, $\langle G^{r},p^{r}
  \rangle$ be two pointed parity games representing $s_{l}$ and
  $s_{r}$ respectively.  Hence $G^{l}$ and $G^{r}$ share the same set
  of positions at infinity $P_{\omega} = X$.  Because of lemmas
  \ref{lemma:normalized} and \ref{lemma:suspensions}, we can assume
  that $\height{G^{l}} =\height{G^{r}} = n$ and that $\kappa(i) = \mu$
  if and only if $i$ is odd for each $i = 1,\ldots ,n$. Given these
  assumptions, we can construct a game $\lcotuple G^{l},
  G^{r}\rcotuple$ of height $n$, having as set of positions the
  disjoint union of the sets $P_{\omega}, P^{l}_{< \omega}, P^{r}_{<
    \omega}$, by pasting together the local structures of $G^{l}$ and
  $G^{r}$.  Recall that, for a pair of functors $F: \Cat{C}\times
  \Cat{E} \rTo \Cat{C}$ and $G : \Cat{D}\times \Cat{E}\rTo \Cat{D}$, a
  pair of initial algebras $(\dgmu{F},\struct{x})$ and
  $(\dgmu{G},\struct{y})$ gives rise to the algebra
  $(\langle\dgmu{F},\dgmu{G}\rangle, \langle
  \struct{x},\struct{y}\rangle)$ of the functor $\langle F\circ
  \prj_{\Cat{C}\times \Cat{E}},G \circ \prj_{\Cat{D}\times
    \Cat{E}}\rangle: \Cat{C} \times \Cat{D} \times \Cat{E}\rTo \Cat{C}
  \times \Cat{D}$, which is moreover an initial one.  Then it is easily
  verified that the relation
  \begin{eqnarray*}
    \val{\lcotuple G^{l}, G^{r}\rcotuple}
    & \iso & \ltuple \val{G^{l}} ,\val{G^{r}}\rtuple
    : \Cat{C}^{P_{\omega}} \rTo 
    \Cat{C}^{P^{l}_{< \omega}} \times \Cat{C}^{P^{r}_{< \omega}}
  \end{eqnarray*}
  holds. In this way we have reduced the problem of finding a
  representation of the $\mu$-functor $\val{s_{l} \land s_{r}}$ to the
  problem of finding a representation of the functor $\prod_{i \in
    \{p^{l},p^{r}\}} \sval{\lcotuple G^{l}, G^{r}\rcotuple}_{i}$ by a
  pointed parity game, which we have previously solved. Figure
  \ref{fig:land} displays the construction of the pointed parity
  game associated to $s_{l} \land s_{r}$.
  \begin{figure}[h] 
    $$
    \mygame{
      []!g{X}="X"
      [d]!O{p_{0}}
      (
      :[ld]
      [d(0.5)]!g{}="LGost1"
      [d]!g{\vdots}="LGost2"
      [d]!g{}="LGost3",
      :[rd]
      [d(0.5)]!g{}="RGost1"
      [d]!g{\vdots}="RGost2"
      [d]!g{}="RGost3"
      )
      "LGost2"[l(0.5)]="StartBackl"
      "RGost2"[r(0.5)]="StartBackr"
      "StartBackl":@`{"StartBackl"+(-1,1),"StartBackl"+(0,3)}"X"
      "StartBackr":@`{"StartBackr"+(1,1),"StartBackr"+(0.5,4)}"X"
      "X"("X"!L{}{\omega})
      "p_{0}"("p_{0}"!L{\mu}{2k+1})
      "LGost1"("LGost1"!L{\nu}{2k})
      "LGost3"("LGost3"!L{\mu}{1})
      "RGost1"("RGost1"!L{\nu}{2k})
      "RGost3"("RGost3"!L{\mu}{1})
    }
    $$
    \vspace{-5mm}
    \caption{Pointed parity game for $s_{l} \land s_{r}$.}
    \label{fig:land}
  \end{figure}  

  Finally, we analyze the case of a term $\mu_{x}.s$. By duality, we
  implicitly analyze the case of a term of the form $\nu_{x}.s$.
  
  Let $\langle G,p_{0} \rangle$ be a parity game such that
  $\sval{G}_{p_{0}} \iso \val{s}$.  Define the game $G' = \langle S' ,
  \hg',\kappa',\epsilon' \rangle$ as follows:
  \begin{itemize}
  \item $S'$ is obtained from  $S$ by adding the move $x
    \rightarrow p_{0}$.
  \item $\hg'(x) = \height{G} + 1$ and $\kappa'(\height{G} +
    1) = \mu$, otherwise $\hg'(p) = \hg(p)$ and $\kappa'(i) =
    \kappa(i)$ if $p \neq x$ and $i \leq \height{G}$.
  \item $\epsilon'(x) = \sigma$ and $\epsilon'(p) =
    \epsilon(p)$ if $p \neq x$.
  \end{itemize}
  Observe that $P(G') = G$ and recall that $\sval{G'}_{x}$ is the
  parameterized initial algebra of the functor in the top composite of
  the diagram below:
  $$
  \mydiagram[6em]{ []( 
    !S 
    {\Cat{C}\times\Cat{C}^{X \setminus
        \{x\}}}
    {\Cat{C}^{P_{<\omega}}\times\Cat{C}^{X}}
    {\Cat{C}^{P_{<\omega}}\times\Cat{C}^{X}}{\Cat{C}} {1}{1.7}, 
    !A {\sval{G}}{ \sval{P(G')}}
    {\prj_{p_{0}}}{\prj(\cod,x)} ) 
    "4":[r]*+{\Cat{C}}^{\coprod}
    "3" :@/^1em/"\Cat{C}"^{\mathcal{E}_{x}} 
  }
  $$
  Thus deduce that $\sval{G'}_{x}$ is also the initial algebra of
  the functor $\coprod \circ \sval{G}_{p_{0}}$. Since this functor is
  naturally isomorphic to $\sval{G}_{p_{0}}$ and therefore to
  $\val{s}$, we obtain the relation $\val{\mu_{x}.s} :=
  \sval{G'}_{x}$.
  \end{proof}

%% file: inset.tex
We recall the game-theoretic interpretation of a parity game $G =
\langle S,\hg,\kappa, \epsilon \rangle$, as a two person game $G(E)$,
parameterized in a choice of sets $E = \{ E_{x} \}_{x \in
  P_{\omega}}$.  The graph $S = \langle P,M,\dom,\cod \rangle$ is a
board with a set of positions $P$ and a set of allowed moves $M$. A
move $m \in M$ is from position $\dom(m)$ to position $\cod(m)$; the
graph $S$ has multiple edges, hence distinct moves relating the same
pair of positions are allowed.  From a position $p$, the set of moves
$M_{p} = \dom^{-1}(p)$ is available, and player $\epsilon(p)$ among
players $\sigma$ and $\pi$ must choose how to move.  If he cannot
move, then he loses.  An infinite play $\gamma_{0} \rightarrow
\gamma_{1}\rightarrow \ldots \gamma_{n} \rightarrow \ldots $ is a win
for player $\sigma$ if and only if $\kappa(\, \max \In \gamma\,) =
\nu$, the set $\In \gamma$ being defined by equation (\ref{eq:in}).
If a play ends in a position $x \in P_{\omega}$, then player $\sigma$
must choose an element $e \in E_{x}$ and then he wins; if $E_{x} =
\emptyset$, then he loses.

A typical element of an inductively defined set is a kind of finite
tree; on the other hand, a typical element of a coinductively defined
set is a kind of infinite tree. We shall see that a similar tree-like
representation is available for parity functors on the category of
sets.

\begin{dfntn}
  Let $\langle S,s_{0} \rangle$ be a pointed graph, a \emph{tree} $T$
  over $\langle S,s_{0} \rangle$ is a non-empty collection of finite
  paths $\gamma$ in $S$ such that $\dom \gamma = s_{0}$, which is
  moreover closed under prefixes: if $\gamma_{1}
  \compp \gamma_{2} \in T$, then $\gamma_{1} \in T$.
\end{dfntn}

Observe that a tree $T$ over $\langle S, s_{0} \rangle$ is itself a
graph if we set $\gamma \rightarrow \gamma'$ if and only if $\gamma' =
\gamma \compp m$ for some $m \in M_{\cod \gamma}$; moreover $\cod: T
\rTo S$ is a morphism of graphs. In particular it makes sense to talk
about an infinite path in $T$.
\begin{dfntn}
  Let $G$ be a parity game and let $E = \{ E_{x} \}_{x \in
    P_{\omega}}$ be a collection of sets. A \emph{deterministic winning
  strategy} for player $\sigma$ from position $p \in P$ in the
  game $G(E)$ is a pair $\langle T, \lambda \rangle$ where $T$ is a
  tree over $\langle S,p \rangle$ with the following properties:
  \begin{itemize}
  \item If $\gamma \in T$, $\epsilon( \cod\gamma ) = \pi$ and $m \in
    M_{\cod \gamma}$, then $\gamma \compp m \in T$.
  \item If $\gamma \in T$ and $\epsilon( \cod\gamma )= \sigma$, then
    there exists a unique $m \in M_{\cod \gamma}$ such that $\gamma
    \compp m \in T$.
  \item Every infinite path in the tree $T$ is a win for player
    $\sigma$, that is, if $\gamma: \hat{\omega} \rTo T$, then
    $\kappa(\, \max \In (\gamma \comp \cod)\,) = \nu$.
  \end{itemize}
  On the other hand, $\lambda$ is a labeling of paths $\gamma \in T$
  such that $\cod \gamma \in P_{\omega}$ by an element $e =
  \lambda(\gamma) \in E_{\cod \gamma}$.  We let $\mathcal{S}_{G,p}(E)$
  be the set of deterministic winning strategies for player $\sigma$
  in the game $G(E)$ from position $p$.
\end{dfntn}

We shall often use $\tree{S},\lab{S}$ for the tree and the label of a
strategy $S$, so that $S = \langle\tree{S},\lab{S} \rangle$.  If
$\langle T,\lambda\rangle, \langle R,\rho\rangle \in
\mathcal{S}_{G,p}(E)$, then we shall write $\langle T,\lambda\rangle
\subseteq \langle R,\rho\rangle$ to mean that $T \subseteq R$ and
$\lambda(\gamma) = \rho(\gamma)$ for all $\gamma \in T$ such that
$\cod \gamma \in P_{ \omega}$.
\begin{lmm}
  \label{lem:subset}
  If $\langle T,\lambda\rangle \subseteq \langle
  R,\rho\rangle$, then $T = R$ and $\lambda =
  \rho$.
\end{lmm}
\begin{proof}
  By induction on the length of $\gamma \in R$. If $\length{\gamma} =
  0$, then $\gamma = 1_{p}$. Since $T$ is non empty, $1_{p} \compp
  \gamma' \in T$, hence $1_{p} \in T$ ($1_{p}$ belongs always to a
  winning strategy from position $p$).  If $\length{\gamma} = n + 1$,
  then we can write $\gamma = \gamma' \compp m$ where
  $\length{\gamma'} = n$.  Since $\gamma' \in R$, by the induction
  hypothesis $\gamma' \in T$ as well. If $\epsilon(\cod \gamma' ) =
  \pi$, then $\gamma' \compp m' \in T$ for each $m' \in M_{\cod
    \gamma'}$, in particular $\gamma = \gamma' \compp m \in T$.  If
  $\epsilon(\cod \gamma') = \sigma$, then there exists $m' \in M_{\cod
    \gamma'}$ such that $\gamma' \compp m' \in T$.  This implies that
  $\gamma' \compp m',\gamma' \compp m \in R$ and $m = m'$, since $R$
  is deterministic. Therefore $\gamma' \compp m \in T$.  Finally, let
  $\gamma \in R$ be such that $\cod \gamma \in P_{\omega}$.  Since we
  have seen that $\gamma \in T$, it follows that $\rho(\gamma) =
  \lambda(\gamma)$.
\end{proof}

Observe that $\mathcal{S}_{G,p}$ is a functor from the category
$\Sets^{P_{\omega}}$ to $\Sets$, the category of sets and functions.
Given a collection of functions $\{ f_{x} : E_{x} \rTo F_{x} \}_{x \in
  P_{\omega}}$, we can transform a strategy $\langle T, \lambda
\rangle \in \mathcal{S}_{G,p}(E)$ into the strategy $\langle T,
f_{\cod}\circ \lambda\rangle \in \mathcal{S}_{G,p}(F)$, where
\begin{eqnarray*}
  (\,f_{\cod}\circ \lambda\,)(\,\gamma\,) & = & f_{\cod \gamma}(\,\lambda(\gamma)\,)\,.
\end{eqnarray*}
Thus, we denote by $\mathcal{S}_{G} : \Sets^{P_{\omega}} \rTo
\Sets^{P_{\leq n}}$ the functor whose $p$-projection is
$\mathcal{S}_{G,p}$, i.e. $\mathcal{S}_{G}(E) = \ltuple
\mathcal{S}_{G,p}(E) \rtuple_{p \in P_{\leq n}}$.  The following
theorem is the main result of this section.
\begin{thrm}
  \label{theo:mainresult}\label{prop:strategies}
  The equality
  \begin{eqnarray*}
    \val{G}(E) & := & 
    \mathcal{S}_{G}(E)
  \end{eqnarray*}
  holds.
\end{thrm}
The equality above means that the collection of sets of deterministic
winning strategies satisfies the universal property involved in the
definition of the parity functor, so that it can be taken to be a
concrete representation of the functor. This equality is reminiscent
of the formula of the Propositional Modal $\mu$-Calculus which
describes the set of winning position for player $\sigma$ in a parity
game, cf. \cite{emersonjutla,BRICS-RS-96-54}.

\begin{figure}[h]
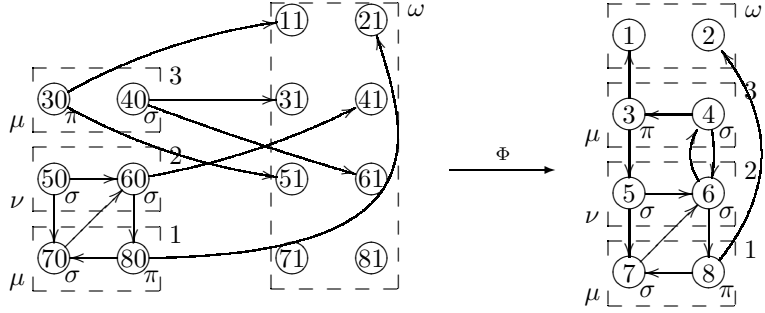

  \centering
  \vspace{-3mm}
  $$    
  \mygame[3em]{
    []
    ([d]!O{30}([r]!P{40})
    [d]!P{50}([r]!P{60})
    [d]!P{70}([r]!O{80})
    )
    [r(3)]!E{11}([r]!E{21}="A")
    [d]!E{31}([r]!E{41})
    [d]!E{51}([r]!E{61})
    [d]!E{71}([r]!E{81})
    "30"(:@/^0.5em/"11",:@/_0.5em/"51")
    "40"(:@/_0.0em/"31",:@/_0.1em/"61")
    "60"(:"80",:@/_0.4em/"41")
    "80"(:"70",:@`{"61"+(1,-1),"41"+(0.5,-0.5)}"21"))
    "50"(:"70",:"60")
    "70"(:"60")
    "11"("81"!L{}{\omega})
    "30"("40"!L{\mu}{3})
    "50"("60"!L{\nu}{2})
    "70"("80"!L{\mu}{1})
  }
  \hspace{-5mm}\rTo[l>=4em]^{\Phi}\hspace{-5mm}
  \mygame[3em]{ 
    []!E{1}([r]!E{2}="A") [d]!O{3}([r]!P{4})
    [d]!P{5}([r]!P{6}) [d]!P{7}([r]!O{8}) "3"(:"1",:"5")
    "5"(:"7",:"6") "7"(:"6") "4"(:"3",:@/^0.2em/"6")
    "6"(:"8",:@/^0.7em/"4") "8"(:"7",:@/_2em/"2") "1"("2"!L{}{\omega})
    "3"("4"!L{\mu}{3}) "5"("6"!L{\nu}{2}) "7"("8"!L{\mu}{1}) 
  }
  $$
  \vspace{-12mm}
  \caption{The delooping game of a parity game.}
  \label{fig:D(L)}
\end{figure}

In the rest of this section we prove theorem \ref{theo:mainresult}.
This is done by induction on the height, observing that it holds in an
obvious way if the height of $G$ is $0$.  Thus we shall suppose in the
following that $\height{G} = n > 0$ and that the statement holds for
the predecessor game $P(G)$. To develop the proof, we shall need a
modified version of the predecessor game that does not completely
forget about the structure in the region of maximal finite height. The
delooping game, displayed on the left in figure \ref{fig:D(L)}, is
devised to detect first passages through the region of maximal finite
height in a parity game.

\begin{dfntn}
  The \emph{delooping game} $D(G) = \langle S',\hg',\kappa',\epsilon'
  \rangle$ of a parity game $G$ is defined as follows:
  \begin{itemize}
  \item $S'$ is the graph whose set of positions is $P_{\leq
      n}\times\{0\} + P\times\{1\}$. For each move $m: p \rightarrow
    p'$ there is a move $(m,0): (p ,0) \rightarrow (p' ,i)$, where $i
    = 1$ if and only if $p' = \cod m \in P_{n} \cup P_{\omega}$ or $p
    = \dom m \in P_{n}$.
  \item $\hg'(p,0) = \hg(p)$ and $\hg(p,1) = \omega$.
  \item $\kappa'(i) = \kappa(i)$ for $i = 1,\ldots ,n = \height{D(G)}=
    \height{G}$.
  \item $\epsilon'(p,0) = \epsilon(p)$.
  \end{itemize}
  The delooping game $D(G)$ comes with a morphism of graphs $\Phi : S'
  \rTo S$, defined by $\Phi(p,i)= p$ and $\Phi(m,0) = m$.
\end{dfntn}

The domain of the functor $\mathcal{S}_{D(G)}$ is the category
$\Sets^{P_{\leq n}\times \{1 \}} \times \Sets^{P_{\omega}\times
  \{1\}}$ and its codomain is the category $\Sets^{P_{\leq n}\times \{
  0 \}}$. Under the obvious isomorphisms $P \iso P \times \{1 \}$ and
$ P_{\leq n} \iso P_{\leq n} \times \{ 0 \}$ we shall look at this
functor as having the shape
$$
\mathcal{S}_{D(G)} : \Sets^{P_{\leq n}} \times \Sets^{P_{\omega}}
\rTo \Sets^{P_{\leq n}}\,.
$$
\begin{lmm}
  There is a natural isomorphism
  \begin{eqnarray*}
    \mathcal{S}_{D(G)}
    &\iso & \langle 
    \mathcal{S}_{P(G)} \circ \prj_{P_{n} \cup P_{\omega}},
    \EQ_{G} \rangle\,.
  \end{eqnarray*}
\end{lmm}
\begin{proof}
  Let $\{ E_{x} \}_{x \in P}$ be a collection of sets.  Observe that
  from a position $(p,0)$ with $p \in P_{< n}$ the game $D(G)(E)$ is
  exactly as the game $P(G)(E')$, $E'$ being the collection $\{ E_{x}
  \}_{x \in P_{n} \cup P_{\omega}}$.  On the other hand, if $p \in
  P_{n}$ then a strategy from position $(p,0)$ in $D(G)(E)$ is given by
  the choice of a tuple $\{ e_{m} \in E_{\cod m}\}_{m \in M_{p}}$
  if $\epsilon(p) = \pi$, or by the choice of a pair $(e,m)$ with $m
  \in M_{p}$ and $e \in E_{\cod m}$ if $\epsilon(p ) = \sigma$.  
\end{proof}

Therefore, by the previous lemma and by the induction hypothesis
$\val{P(G)}
\linebreak
\iso \mathcal{S}_{P(G)}$, in order to prove theorem
\ref{theo:mainresult} it is enough to show that the collection of sets
$\mathcal{S}_{G}(E)$ carries an invertible algebra structure for the
functor $\mathcal{S}_{D(G)}(-,E)$, and that this structure leads to an
initial algebra if $\kappa(n) = \mu$ and to a final coalgebra if
$\kappa(m) = \nu$.

Observe that an infinite play $\delta$ is a win for player $\sigma$
in $D(G)$ if and only if $\Phi \delta$ is an infinite winning play for
player $\sigma$ in $G$. Then, it is informally seen that a winning
strategy $\langle R,\rho \rangle$ for player $\sigma$ from a position
$(p,0)$ in $D(G)(\mathcal{S}_{G}(E),E)$ gives rise to a strategy from
position $p$ in $G(E)$ as follows: player $\sigma$ uses $\langle R,
\rho \rangle$ as far as he can, by identifying a position $p$ to the
position $(p,0)$; if this is not anymore possible, since a position of
the form $(p',1)$ with $p' \in P_{\leq n}$ has been reached by means of
a play $\delta \in R$, then player $\sigma$ continues according to the
strategy $\rho(\delta) \in \mathcal{S}_{G,p'}(E)$. Moreover, every
winning strategy from $p$ in $G(E)$ arises in this way. We formalize
these ideas next.

\begin{dfntn} \hspace{2mm}
  \begin{itemize}
  \item A path $\delta$ of $D(G)$ is called an \emph{atom} if
    $\length{\delta} > 0$, $\cod \delta \in P'_{\omega}$ and $\cod
    \Phi\delta \in P_{\leq n}$. If $R$ is a set of paths in $D(G)$,
    then we shall write $A(R)$ for the set of atoms of $R$.
  \item Let $\delta$ be a path of $G$ and let $T$ be a collection of
    paths of $G$ with domain $\cod \delta$. By $\delta \compp T$ we
    mean the set $\{ \,\delta \compp \gamma \, |\, \gamma \in T \,\}$.
  \item Let $\langle T ,\lambda \rangle \in \mathcal{S}_{G,p}(E) $ and
    $\delta \in T$.  By $\delta \res \langle T ,\lambda \rangle$ we
    denote the winning strategy $\langle T', \lambda' \rangle \in
    \mathcal{S}_{G, \cod \delta}(E)$ where $T'$ is the set $\{
    \,\gamma\,| \,\delta \compp \gamma \in T \,\}$ and
    $\lambda'(\gamma) = \lambda( \delta \compp \gamma )$ if $\gamma
    \in T'$ and $\cod \gamma \in P_{\omega}$. We call this strategy
    \emph{the residual strategy} of $\langle T,\lambda\rangle$ after
    the path $\delta$.
\end{itemize}
\end{dfntn}

\begin{lmm}
  The pair $\langle T' ,\lambda' \rangle = \delta \res \langle
  T,\lambda\rangle$ is a winning strategy for player $\sigma$ in
  the game $G(E)$ from position $\cod \delta$.
\end{lmm}
\begin{proof}
  Let $\gamma \in T'$. If $\cod \gamma \in P_{\leq n}$, then
  $\cod(\delta \compp \gamma) = \cod \gamma \in P_{\leq n}$ as
  well.  If $\epsilon(\cod \gamma) = \pi$, then $\delta \compp
  \gamma \compp m \in T$ and $\gamma \compp m \in T'$ for all $m
  \in M_{\cod \gamma}$. If $\epsilon(\cod \gamma) = \sigma$, then
  there exists an $m \in M_{\cod \gamma}$ such that $\delta \compp
  \gamma \compp m \in T$, and henceforth $\gamma \compp m \in T'$.
  If $\gamma \compp m' \in T'$, then $\delta \compp \gamma \compp
  m' \in T$ and therefore $m = m'$.  Let $\gamma$ be an infinite
  path in $T'$, then $\delta \compp \gamma$ is an infinite path in
  $T$ and since $ \In( \gamma ) = \In( \delta \compp\gamma )$ we
  see that
  \begin{eqnarray*}
    \kappa(\max \In( \gamma )) 
    & = & \kappa(\max\In( \delta \compp\gamma )) \\
    & = & \nu\,.
  \end{eqnarray*}
\end{proof}

\begin{dfntn}
  We define an algebra
  $$
  \struct{x} : 
  \mathcal{S}_{D(G)}(\mathcal{S}_{G}(E),E)
  \rTo 
  \mathcal{S}_{G}(E)
  $$
  in the following way. If $\langle R, \rho \rangle \in
  \mathcal{S}_{D(G)}(\mathcal{S}_{G}(E),E)$, then we let
  $\struct{x}_{p}(R,\rho) = \langle \low{R},\low{\rho} \rangle$ where
  \begin{eqnarray*}
    \low{R} & = & 
    \Phi R \cup 
    \bigcup_{\delta \in A(R)} 
    \Phi\delta \compp \tree{\rho(\delta)}\,, \\
    \low{\rho}(\gamma)
    & = & \left \{ 
      \begin{array}{ll}
        \rho(\delta), & \gamma = \Phi\delta,\, \delta\in R, \\
        \lab{\rho(\delta)}(\gamma'), 
        & \gamma = (\Phi\delta) 
        \compp \gamma', \,\delta \in A(R),
      \end{array}
    \right .
  \end{eqnarray*}
  if $\gamma \in \low{R}$ and $\cod \gamma \in P_{\omega}$.
\end{dfntn}
\begin{lmm}
  $\chi_{p}( R, \rho )$ is a winning strategy for player $\sigma$
  in the game $G(E)$ from position $p$.
\end{lmm}
\begin{proof}
  Let $\gamma \in \low{R}$ be such that $\cod \gamma \in P_{\leq
    n}$. Either $\gamma =\Phi\delta$ where $\delta \in R$ is not
  an atom, or $\gamma = \Phi\delta \compp \gamma'$ where $\delta
  \in A(R)$ and $\gamma' \in \tree{\rho(\delta)}$.  Similarly, an
  infinite path $\gamma$ in $\low{R}$ is either the image of
  infinite path in $R$ or it is of the form $\gamma = \Phi\delta
  \compp \gamma'$ where $\delta \in A(R)$ and $\gamma'$ is an
  infinite path in $\tree{\rho(\delta)}$. The desired properties of
  $\low{R}$ follow then from the properties of $R$ and from the
  properties of $\rho(\delta)$, respectively. For example, let
  $\gamma =\Phi\delta$ where $\delta \in R$ is not an atom.
  Suppose that $\epsilon(\cod \Phi\delta) = \sigma$ and observe
  that $\epsilon(\cod \delta) = \sigma$ as well, since $\delta$ is
  not an atom. Hence we can find $m \in M_{\cod \delta}$ such that
  $\delta \compp m \in R$ and therefore $\Phi\delta \compp \Phi m
  \in \low{R}$. If $\Phi\delta \compp m' \in \low{R}$, then we can
  find $\delta'$ such that $\Phi \delta' = \Phi\delta \compp m'$.
  It follows that $\delta' = \delta \compp m''$, where $\Phi m'' =
  m'$. Since $R$ is deterministic, $m'' = m$ and hence $m' = \Phi m$.
\end{proof}   
 
\newpage
\begin{dfntn}
  We define a coalgebra
  $$
  \struct{y} : 
  \mathcal{S}_{G}(E)\rTo
  \mathcal{S}_{D(G)}(\mathcal{S}_{G}(E),E)
  $$
  in the following way. If $\langle T, \lambda \rangle
  \in\mathcal{S}_{G}(E)$, then we let $\struct{y}_{p}(T,\lambda)
  = \langle \rise{T},\rise{\lambda} \rangle$ where
  \begin{eqnarray*}
    \rise{T} & = & \{ \, \delta \,|\, \Phi\delta \in T\, \} \\
    \rise{\lambda}(\delta) & = & 
    \left \{
      \begin{array}{ll}
        \lambda(\Phi\delta),& \cod \Phi\delta \in P_{\omega}, \\
        \Phi\delta \res \langle T,\lambda\rangle, 
        & \cod \Phi\delta \in P_{\leq n},
      \end{array}
    \right .
  \end{eqnarray*}
  if $\delta \in \rise{T}$ and $\cod \delta \in P'_{\omega}$.
\end{dfntn}

\begin{lmm}
  $\struct{y}_{p}(T,\lambda)$ is a winning strategy for player
  $\sigma$ from position $p$ in the game $D(G)(\mathcal{S}_{G}(E),E)$.
\end{lmm}
\begin{proof}
  Let $\delta \in \rise{T}$ and suppose that $m \in M_{\cod \delta}$
  and $\epsilon(\cod \delta) = \pi$. Then $\epsilon(\cod \Phi \delta)
  = \pi$, so that $\Phi\delta \compp \Phi m \in T$ and thus $\delta
  \compp m \in \rise{T}$. If $\epsilon(\cod \delta) = \sigma$, then
  $\delta$ is not an atom.  Since $\epsilon(\cod \Phi \delta) =
  \sigma$, there exists an $m$ such that $\Phi \delta \compp m \in T$.
  Since $\delta$ is not an atom, the move $m$ can be lifted to a move
  $m'\in M_{\cod \delta}$ such that $\Phi m' = m$. It follows that
  $\delta \compp m' \in \rise{T}$.  If $\delta \compp m'' \in
  \rise{T}$, then $\Phi m'' = m$, since $T$ is deterministic, and
  therefore $m'' = m'$ since a lifting of $m$ is unique. Finally, an
  infinite path $\delta$ in $\rise{T}$ gives rise to an infinite path
  $\Phi \delta$ in $T$, which is a win for player $\sigma$ in the game
  $G$.  From this it follows that $\delta$ is win for player $\sigma$
  in $D(G)$.
\end{proof}

\begin{prpstn}
  The functions $\struct{x}_{p}$ and $\struct{y}_{p}$ are
  inverse to each other.
\end{prpstn}
\begin{proof}
  Let $\langle R,\rho\rangle \in
  \mathcal{S}_{D(G),p}(\mathcal{S}_{G}(E),E)$ and $\langle
  T,\lambda \rangle \in \mathcal{S}_{G,p}(E)$.  We shall
  show that
  \begin{eqnarray*}
    \langle R,\rho\rangle \subseteq 
    \struct{y}_{p}(T,\lambda)
    & \tiff &
    \struct{x}_{p}(R,\rho)
    \subseteq \langle T,\lambda\rangle\,.
  \end{eqnarray*}
  The desired result will follow from lemma \ref{lem:subset}.  
  
  Suppose first that $\langle R,\rho\rangle \subseteq
  \struct{y}_{p}(T,\lambda)$. Thus $R \subseteq \Phi^{-1}T$ and $\Phi R
  \subseteq T$, and if $\delta \in R$ is an atom, then
  $\rho(\delta) = \rise{\lambda}(\delta) = \Phi\delta \res \langle
  T,\lambda \rangle$. Hence
  \begin{eqnarray*}
    \low{R} & = & 
    \Phi R \cup \bigcup_{\delta \in A(R)}
    \Phi\delta \compp 
    \tree{\Phi\delta \res \langle T, \lambda \rangle } \\
    & \subseteq & T\,.
  \end{eqnarray*}
  Consider a path $\gamma \in \low{R}$ such that $\cod \gamma \in
  P_{\omega}$. If $\gamma = \Phi\delta$, then $\low{\rho}(\gamma) =
  \rho(\delta) = \rise{\lambda}(\delta) = \lambda(\Phi\delta)$, 
  and if $\gamma =
  \Phi\delta \compp \gamma'$, then $\low{\rho}(\gamma) =
  \lab{\rho(\delta)}(\gamma') =  \lab{\Phi\delta \res \langle
  T,\lambda \rangle}(\gamma') = \lambda(\Phi\delta\compp
  \gamma') = \lambda(\gamma)$.
  
  Suppose now that $\struct{x}_{p}(R,\rho) \subseteq \langle
  T,\lambda\rangle$. Then $\Phi R \subseteq T$ and therefore $R
  \subseteq \Phi^{-1}R = \rise{T}$. Consider a path $\delta \in R$
  such that $\cod \delta \in P'_{\omega}$. If $\cod \Phi\delta \in
  P_{\omega}$ then $\rise{\lambda}(\delta) = \lambda(\Phi\delta) =
  \low{\rho}(\Phi\delta) = \rho(\delta)$; on the other hand, if $\cod
  \Phi\delta \in P_{\leq n}$, then $\rise{\lambda}(\delta) =
  \Phi\delta \res \langle T,\lambda\rangle = \rho(\delta)$, since
  $\rho(\delta) \subseteq \Phi\delta \res \langle T, \lambda \rangle$.
  This can be seen as follows: if $\gamma \in \tree{\rho(\delta)}$ then
  $\Phi\delta \compp \gamma \in \low{R} \subseteq T$, so that $ \gamma
  \in \tree{\Phi\delta \res \langle T,\lambda\rangle}$; if moreover
  $\cod \gamma \in P_{\omega}$, then $\lab{\rho(\delta)}(\gamma) =
  \low{\rho}(\Phi\delta \compp \gamma) = \lambda(\Phi\delta \compp
  \gamma) = \lab{\Phi\delta \res \langle T,\lambda\rangle}(\gamma)$.
\end{proof}

\begin{prpstn} 
  \label{prop:initialalgebra}
  If  $\kappa(n) = \mu$, then
  $$
  \struct{x} : 
  \mathcal{S}_{D(G)}(\mathcal{S}_{G}(E),E) \rTo \mathcal{S}_{G}(E)
  $$
  is an initial $\mathcal{S}_{D(G)}(-,E)$-algebra.
\end{prpstn}
\begin{proof}
  First we construct a graph $\mathcal{G}$ as follows: its vertices
  are pairs $(S,p)$ with $p \in P_{\leq n}$ and $S \in
  \mathcal{S}_{G,p}(E)$. A transition of this graph is of the form
  $$
  \delta : (S,p) \rightarrow (\Phi\delta \res S,p')
  $$
  where $\delta$ is an atom such that $\Phi\delta \in S$, $\dom
  \Phi\delta = p$ and $\cod \Phi\delta = p'$. Observe that the graph
  $\mathcal{G}$ is well founded: given an infinite sequence
  $\delta_{i} : (S_{i-1},p_{i-1}) \rightarrow (S_{i},p_{i})$, $i \geq
  1$, we can construct the infinite path $\Phi\delta_{1} \compp
  \Phi\delta_{2}\compp \ldots $ which belongs to $S_{0}$ and
  contradicts the condition on infinite paths for a winning strategy.
  Observe moreover that if $\langle \rise{T},\rise{\lambda}\rangle =
  \struct{y}_{p}(T,\lambda)$ and $\delta \in A(\rise{T})$, then
  $\rise{\lambda}(\delta) = \Phi\delta \res \langle T,\lambda\rangle$,
  hence $\delta : (\langle T,\lambda\rangle,p) \rightarrow \langle
  \rise{\lambda}(\Phi\delta), \cod \Phi\delta\rangle$.
  
  Thus, if $\beta : \mathcal{S}_{D(G)}(B,E)\rTo B$ is another algebra,
  then we can define $f : \mathcal{S}_{G}(E) \rTo B$, by the formula:
    \begin{eqnarray*}
      f_{p}(T,\lambda)
      & = & 
      \beta_{p}(\,\mathcal{\mathcal{S}}_{D(G),p}(f,E)(\,\struct{y}_{p}(T,\lambda)\,)\,)\ \\
      & = & 
      \beta_{p}(\,\mathcal{\mathcal{S}}_{D(G),p}(f,E)(\,\rise{T},\rise{\lambda}\,)\,)\\
      & = & 
      \beta_{p}(\,\rise{T},\lambda'\,)
    \end{eqnarray*} 
    where 
    \begin{eqnarray*}
      \lambda'(\delta)
      & = & 
      \left \{
        \begin{array}{ll}
          \lambda(\delta) , & \cod \Phi\delta \in P_{\omega} \\
          f_{\cod \Phi \delta}(\rise{\lambda}(\delta)), 
          & \delta \textrm{ an atom}
        \end{array}
      \right .
      \end{eqnarray*}
      using the induction hypothesis (on the well founded graph
      $\mathcal{G}$) that we have previously defined $f_{p'}(S')$ for
      each pair $(S',p')$ such that $(\langle T,\lambda \rangle,p)
      \rightarrow (S',p')$.  This is also the unique way to define $f$
      so that $\struct{x} \comp f = \mathcal{S}_{D(G)}(f,E) \comp
      \beta$.  
  \end{proof}
  
  \begin{prpstn}
    \label{prop:finalcolagebra}
    If  $\kappa(n) = \nu$, then
    $$
    \struct{y} 
    : \mathcal{S}_{G}(E) \rTo \mathcal{S}_{D(G)}(\mathcal{S}_{G}(E),E) 
    $$
    is a final $\mathcal{S}_{D(G)}(-,E)$-coalgebra.
  \end{prpstn}
  \begin{proof}
    Consider a coalgebra 
    $$
    \beta : B \rTo
    \mathcal{S}_{D(G)}(B,E)\,,
    $$
    we first define a graph $\mathcal{G}_{\beta}$ as follows: a
    state of $\mathcal{G}_{\beta}$ is a pair $(b,p)$ such that $p \in
    P_{\leq n}$ and $b \in B_{p}$, and a transition $(b,p) \rightarrow
    (b',p')$ of $\mathcal{G}_{\beta}$ is an atom $\delta \in
    \tree{\beta_{p}(b)}$ such that $\cod \Phi\delta = p'$ and
    $\lab{\beta_{p}(b)}(\delta) = b'$. Observe that in the proof of
    proposition \ref{prop:initialalgebra} the graph $\mathcal{G}$
    coincides with the graph $\mathcal{G}_{\struct{y}}$ defined here.
    
    We now define a collection of functions $\{ f_{p} : B_{p} \rTo
    \mathcal{S}_{G,p}(E) \}_{ p \in P_{\leq n}}$ and then split the
    proof of proposition \ref{prop:finalcolagebra} in a sequence of
    lemmas: in \ref{lem:welldefined} we show that these functions are
    well defined, in \ref{lem:morphism} we show that this defines a
    morphism of coalgebras, and finally in \ref{lem:unique} we show
    that this is the unique such morphism.
    \begin{dfntn}
      For each $p \in P_{\leq n}$ and $b \in B_{p}$ we define
      $f_{p}(b) = \langle T_{b}, \lambda_{b} \rangle \in
      \mathcal{S}_{G,p}(E)$ as follows.  We say that $\gamma \in
      T_{b}$ if and only if $\gamma$ has a factorization of the form
      \begin{eqnarray}
        \label{eq:factorization}
        \gamma  & = & 
        \Phi\delta_{1} \compp \ldots 
        \compp \Phi\delta_{k} \compp \Phi\delta_{k
          + 1}
      \end{eqnarray}
      such that
      \begin{itemize}
      \item[1.] $\Delta = (\delta_{1},\ldots ,\delta_{k})$ is a path in
        $\mathcal{G}_{\beta}$ such that $\dom \Delta = (b,p)$ and
        $\cod \Delta = (b',p')$,
      \item[2.] $\delta_{k + 1} \in \tree{\beta_{p'}(b')}$ is not an
        atom.
      \end{itemize}
      If $\cod \gamma \in P_{\omega}$, then we let
      $\lambda_{b}(\gamma) = \lab{\beta_{p'}(b')}(\delta_{k + 1})$.
    \end{dfntn}
    We remark that a factorization of the form
    (\ref{eq:factorization}), without the additional requirements $1.$
    and $2.$, exists for any path $\gamma$ and is unique. This follows
    from the observation that the set of paths
    $\{\,\Phi\delta\,|\,\textrm{$\delta$ is an atom}\,\}$ does not
    contain comparable elements with respect to the prefix order.
    Using the language of the theory of codes \cite{reutenauer}, this
    set is a prefix code. Recall also that $\delta_{k + 1}$ is not an
    atom if either $\length{\delta_{k + 1}} = 0$ or $\cod \delta_{k +
      1} \in P'_{\omega}$ implies $\cod \Phi\delta_{k + 1} \in
    P_{\omega}$.
  
  The game-theoretic interpretation of the strategy $f_{p}(b)$ is as
  follows.  From position $p$, player $\sigma$ uses the strategy
  $\beta_{p}(b)$ as long as he can.  As soon as the play reaches a
  position $p'$ such that either $p'$ in $P_{n}$ or after one move if
  $p \in P_{n}$, this strategy becomes unavailable.  However, if one
  of these two cases happens, the strategy $\beta_{p}(b)$ gives player
  $\sigma$ the choice of an element $b' = \lab{\beta_{p}(b)}(\delta)$
  in $B_{p'}$ and therefore the choice of a new strategy
  $\beta_{p'}(b')$.  Thus player $\sigma$ iterates this process.
  Iteration of this process is expressed by saying that the residual
  strategy of $f_{p}(b)$ after the image of an atom $\delta$ is the
  strategy $f_{p'}(b')$. This is the content of the next lemma.
  \begin{lmm}
    \label{lemma:iteration}
    Let $\langle R, \rho \rangle = \beta_{p}(b)$ and let $\delta
    \in A(R)$. Then
    \begin{eqnarray*}
      \Phi\delta \res f_{p}(b) & =  & f_{\cod
        \Phi\delta}( \rho(\delta)) \,.
    \end{eqnarray*}
  \end{lmm}
  \begin{proof}
    Let $\gamma \in \tree{f_{\cod \Phi\delta}(\rho(\delta))}$, then we
    can write $\gamma = \Phi\delta_{1} \compp \ldots \compp \Phi
    \delta_{k+1}$ and thus $\Phi\delta \compp \gamma = \Phi\delta
    \compp\Phi\delta_{1} \compp \ldots\compp \Phi \delta_{k+1}$ shows
    that $\Phi\delta \compp \gamma \in \tree{f_{p}(b)}$ and $\gamma
    \in \tree{\Phi\delta \res f_{p}(b)}$. If $\cod \gamma \in
    P_{\omega}$ then $\lab{\Phi\delta \res f_{p}(b)}(\gamma) =
    \lab{f_{p}(b)}(\Phi\delta \compp \gamma) =
    \lab{\beta_{p'}(b')}(\delta_{k + 1})$ and similarly $\lab{f_{\cod
        \Phi\delta}(\rho(\delta))}( \gamma) =
    \lab{\beta_{p'}(b')}(\delta_{k + 1})$. 
  \end{proof}
  \begin{lmm}
    \label{lem:welldefined}
    The pair $f_{p}(b) = \langle T_{b},\lambda_{b} \rangle$ is a
    winning strategy for player $\sigma$ in the game $G(E)$ from
    position $p$.
  \end{lmm}
  \begin{proof}
    Let $\gamma \in T_{b}$ have a factorization $\Phi\delta_{1}\compp
    \ldots \compp \Phi\delta_{k+1}$. Observe that $\delta_{1} \in
    \beta_{p}(b)\in \mathcal{S}_{D(G),p}(B,E)$ implies that $\dom
    \delta_{1} = (p,0)$, hence $\dom \gamma = \dom \Phi \delta_{1} =
    p$.  From the definition it is clear that $T_{b}$ is closed under
    prefixes.
    
    Let $\gamma \in T_{b}$ and suppose first that $\epsilon(\cod
    \gamma) = \pi$. If $m \in M_{\cod \gamma}$, then $(m,0) \in
    M_{\cod \delta_{k + 1}}$, since $\delta_{k + 1}$ is not an atom,
    hence $ \delta'_{k + 1} = \delta_{k+1} \compp m \in
    \beta_{p'}(b')$. If $\delta'_{k + 1}$ is not an atom, then we can
    write
    \begin{eqnarray*}
      \gamma  \compp m& = & 
      \Phi\delta_{1} \compp \ldots \compp \Phi\delta_{k} 
      \compp \Phi\delta'_{k
        + 1}
    \end{eqnarray*}
    and if $\delta'_{k + 1}$ is an atom we can write
    \begin{eqnarray*}
      \gamma  \compp m& = & 
      \Phi\delta_{1} \compp \ldots \compp \Phi\delta_{k} 
      \compp \Phi\delta'_{k
        + 1} \compp \Phi 1_{(\cod \Phi \delta'_{k + 1},0)}\,,
    \end{eqnarray*}
    If we let $p'' = \cod \Phi \delta_{k + 1}'$ and $b'' =
    \lab{\beta_{p'}(b')}(\delta_{k +1}')$, then we observe $1_{
      (\cod \delta'_{k + 1},0)} = 1_{(p'',0)}\in
    \tree{\beta_{p''}(b'')}$.  In both cases we conclude that
    $\gamma \compp m \in T_{b}$.
    
    If $\epsilon(\cod \gamma) = \sigma$, then $\epsilon'(\cod
    \delta_{k + 1}) = \sigma$ so that $\delta_{k + 1} \compp (m,0) \in
    \beta_{p'}(b')$ for a unique $m \in M_{\cod \gamma}$. As before,
    we conclude that $\gamma \compp m \in T_{b}$. On the other hand,
    if $\gamma \compp m' \in T_{b}$, then $\delta_{k + 1} \compp
    (m',0) \in \beta_{p'}(b')$, since such a factorization for
    $\gamma$ is unique. Thus $m = m'$, since $\beta_{p'}(b')$ is
    deterministic.
    
    Consider now an infinite path $\gamma$ in $T_{b}$.  Either this
    infinite path visits the region $P_{n}$ infinitely often, in which
    case it is a win for player $\sigma$, or we can write $\gamma =
    \gamma'\compp \Phi\delta$, where $\delta$ is an infinite play in
    $D(G)(B,E)$, played according to a given winning strategy for this
    game.  This infinite play is a win for player $\sigma$ in
    $D(G)(B,E)$ which implies that $\gamma$ is a win for player
    $\sigma$ in $G(E)$.
  \end{proof}
  
  \begin{lmm}
    \label{lem:morphism}
    The  diagram
    $$
    \mydiagram[6em]{ [](!S{B}
      {\mathcal{S}_{D(G)}(B,E)}{\mathcal{S}_{G}(E)}
      {\mathcal{S}_{D(G)}(\mathcal{S}_{G}(E),E)}{1}{2},
      !a{^{\beta}}{^{f}}{^/-0.6em/{\mathcal{S}_{D(G)}(f,E)}}{^{\struct{y}}}
    }
    $$
    commutes.
  \end{lmm}
  \begin{proof}
    It is enough to show that for all $p \in P_{\leq n}$ and $b \in
    B_{p}$
    \begin{eqnarray*}
      \mathcal{S}_{D(G),p}(f,E)(\, \beta_{p}(b)\,)
      & \subseteq & \struct{y}_{p}(\,f_{p}(b)\,)\,.
    \end{eqnarray*}
    Let $\langle T_{b},\lambda_{b}\rangle = f_{p}(b)$ and $\langle R,
    \rho\rangle = \beta_{p}(b)$. If $\delta \in R$, then $\Phi\delta
    \in T_{b}$ so that $\delta \in \rise{T_{b}}$.  Suppose now that
    $\cod \delta \in P'_{\omega}$. If $\cod \Phi\delta \in
    P_{\omega}$, then $\rise{ \lambda_{b}}(\delta)= \lambda_{b}(\Phi
    \delta) = \rho(\delta)$.  If $\cod \Phi \delta \in P_{\leq n}$,
    that is, if $\delta$ is an atom, then $\rise{\lambda_{b}}(\delta)
    = \Phi\delta \res f_{p}(b) = f_{\cod\Phi \delta}( \rho(\delta))$,
    by lemma \ref{lemma:iteration}.
  \end{proof}
  
  \begin{lmm}
    \label{lem:unique}
    If a collection of functions $g: B \rTo \mathcal{S}_{G}(E)$
    satisfies the relation $g \comp \struct{y} = \beta \comp
    \mathcal{S}_{D(G)}(g,E)$, then $g = f$.
  \end{lmm}
  \begin{proof}
    We will prove that $g_{p}(b) \subseteq f_{p}(b)$ for all $p \in
    P_{\leq n}$ and $b \in B_{p}$. In the following let $\beta_{p}(b)
    = \langle R ,\rho\rangle$ and recall that
    $\mathcal{S}_{D(G)}(g,E)(R,\rho) = \langle R ,\rho' \rangle$ where
    $\rho'(\delta) = \rho(\delta)$ if $\cod \Phi \delta \in
    P_{\omega}$ and $\rho'(\delta) = g_{\cod \Phi \delta}(
    \rho(\delta))$ if $\delta$ is an atom. Thus we have reduced the
    relation $g_{p}(b) = \struct{x}_{p}(\,\mathcal{S}_{D(G),p}(g,E)(
    \beta_{p}(b))\,)$ to the relation $g_{p}(b) =
    \struct{x}_{p}(R,\rho')$.
    
    As a first part, we  prove by induction on the length of
    $\gamma$ the following statement: \emph{for each $p \in P_{\leq
        n}$ and $b \in B_{p}$, if $\gamma \in \tree{g_{p}(b)}$,
      then $\gamma \in \tree{f_{p}(b)}$}.
    
    The statement is trivial if $\length{\gamma} = 0$, since if
    $\gamma \in \tree{g_{p}(b)}$, then $\gamma = 1_{p}$ and $1_{p}$
    belongs to any winning strategy from position $p$.  If
    $\length{\gamma} > 0$, we argue using the equality $g_{p}(b) =
    \struct{x}_{p}(R,\rho')$. If $\gamma = \Phi\delta$, then $\gamma
    \in T_{b}$ by its definition. If $\gamma = \Phi\delta \compp
    \gamma'$, where $\delta$ is an atom of $R$ and $\gamma ' \in
    \tree{\rho'(\delta)} = \tree{g_{\cod \Phi\delta}(\rho(\delta))}$,
    then $\gamma' \in \tree{f_{\cod \delta}(\rho(\delta))}$, since
    $\length{\gamma '}< \length{\gamma}$ and using the induction
    hypothesis. Then it is easily seen that $\gamma = \Phi\delta \compp
    \gamma' \in \tree{f_{p}(b)}$ as well.
    
    We now prove again by induction on the length the following
    statement: \emph{for each $p \in P_{\leq n}$ and $b \in B_{p}$, if
      $\gamma \in g_{p}(b)$ and $\cod \gamma \in P_{\omega}$, then
      $\lab{g_{p}(b)}(\gamma) = \lab{f_{p}(b)}(\gamma)$}.
    
    The statement is again obvious if $\length{\gamma} = 0$, since
    there is no such $\gamma$ with $\dom \gamma \in P_{\leq n}$ and
    $\cod \gamma \in P_{\omega}$.  If
    $\length{\gamma} > 0$, then two cases. Either $\gamma = \Phi\delta$ with $\delta
    \in R$, in which case $\lab{g_{p}(b)}(\gamma) =
    \low{\rho'}(\Phi\delta) = \rho'(\delta) = \rho(\delta) =
    \lambda_{b}(\gamma)$ by the definition of $f$.  Or $\gamma =
    \Phi\delta \compp \gamma'$ where $\delta$ is an atom of $R$ and
    $\gamma' \in \tree{\rho'(\delta)} = \tree{g_{\cod \delta}(
      \rho(\delta))}$.  In this case
    $$
    \renewcommand{\arraystretch}{1.3}
    \begin{array}[b]{rcl@{\hspace{7mm}}l}
      \lab{g_{p}(b)}(\gamma) 
      & = &  \low{\rho'}(\Phi\delta \compp \gamma')
      & g_{p}(b) = \struct{x}_{p}(R,\rho')\\
      & = & \lab{\rho'(\delta)}(\gamma') 
      & \textrm{def. of $\struct{x}_{p}$}\\
      & = &
      \lab{g_{\cod
          \Phi\delta}( \rho(\delta))}(\gamma')
      & \textrm{def. of $\rho'$} \\
      & = & \lab{f_{\cod
          \Phi\delta}(\rho(\delta))}(\gamma')  
      & \textrm{induction hypothesis on $\gamma'$}\\
      & = & \lambda_{b}(\Phi\delta \compp \gamma') 
      & \Phi\delta \res f_{p}(b) 
      = f_{\cod \Phi\delta}(\rho(\delta))\\
      & = & \lambda_{b}(\gamma)\,.
    \end{array}
    \renewcommand{\arraystretch}{1}
    $$
  \end{proof}
  This ends the proof of proposition \ref{prop:finalcolagebra} too.
\end{proof}
  
Thus we have completed the proof of theorem \ref{theo:mainresult}. We
end this section with some examples illustrating the theory so far
developed.

\newcommand{\cons}{\textrm{cons}}
\newcommand{\nil}{\textrm{nil}}

\begin{xmpl}
  We consider the set of finite lists over a set of symbols $E$. This
  is initial algebra of the functor $1 + (Y \times E)$ and therefore
  it is the denotation of the $\mu$-term $\mu_{y}.(\top \vee (y \land
  E))$.
  \begin{figure}[h]
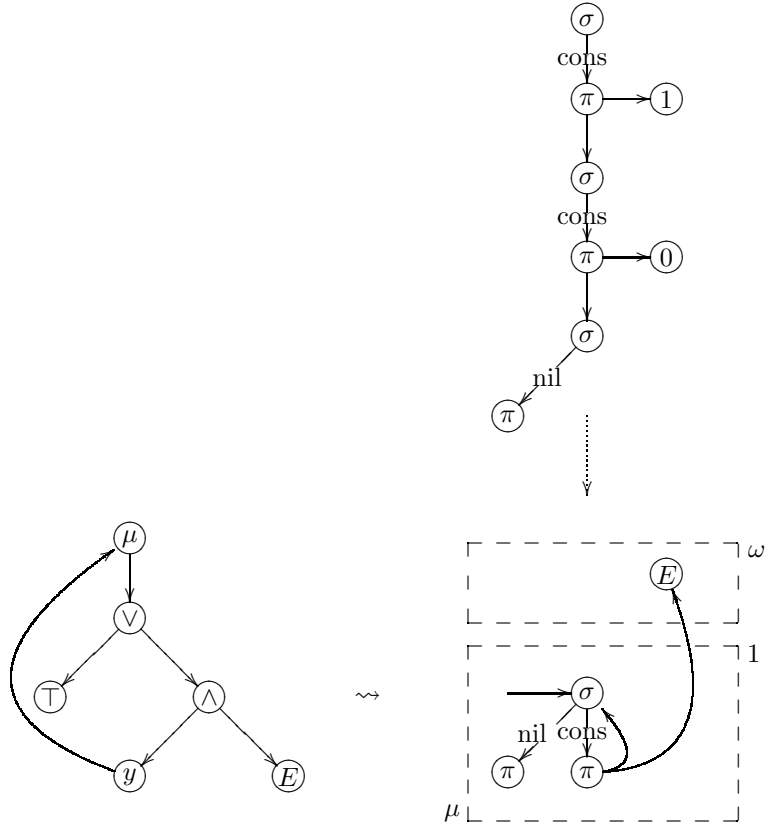

    $$
    \mygame[3em]{
      [d(6)] []!E{\mu} 
      :[d]!E{\vee} 
      ( :[ld]!E{\top}
      [ll]="I1", 
      :[rd]!E{\land} 
      ( :[ld]!E{y}:@`{"I1"}"\mu",
      :[rd]!E{E}
      ) 
      "\land"[r(2)]*+{\leadsto} 
    } 
    \hspace{-0mm} 
    \mygame[3em]{
      []="S"([r]!E{E}) [d(1.5)] [l] :[r]!E{\sigma}="A" (
      :[ld]!E{\pi}|{\makebox[8mm][l]{\nil}} [rr]="I1",
      [rrd]="I2", :[d]!E{\pi}="B"|{\makebox[8mm][l]{\textrm{cons}}} (
      :@`{"I1"}"\sigma", :@`{"I2"}"E" ) "S"
      [l(1.5)u(0.4)]("S"[r(1.5)d(0.2)]!L{}{\omega}) "A"
      [l(1.5)u(0.6)]("B"[r(1.5)d(0.2)]!L{\mu}{1}) "S" [u(7)]
      !E{\sigma} :[d]!E{\pi}|{\makebox[8mm][l]{\cons}}
      (:[r]!E{1}) :[d] !E{\sigma}
      :[d]!E{\pi}|{\makebox[8mm][l]{\cons}} (:[r]!E{0}) :[d]
      !E{\sigma} (:[dl]!E{\pi}|{\makebox[4mm][l]{\nil}})
      [d]:@{{.}{.}{>}}[d] }
    $$\centering\vspace{-8mm}
    \caption{Lists as winning strategies.}
    \label{fig:lists}
  \end{figure}
  In figure \ref{fig:lists} we have translated this $\mu$-term
  into a pointed parity game, according to proposition
  \ref{prop:termstogames} and to a well established practice in the
  model checking community.  The conventions are the ones followed
  until now: positions of the games, labeled by $\sigma$ or $\pi$, are
  grouped within boxes according to their height.  The height is on
  the right of the boxes, the color is on the left. For convenience of
  exposition, we have labeled transitions in the figure, even if this
  is not strictly necessary.
  
  It is immediate to realize that there is a bijection between lists
  and deterministic winning strategies in the parity game. If we let
  $E = \{ 0,1 \}$, we have represented in figure \ref{fig:lists} the
  list $\cons(\cons(\nil,0),1)$ in the form of a winning strategy, the
  tree over the game.  Observe that we cannot obtain infinite lists
  since every infinite path on the corresponding tree would be a loss
  for player $\sigma$.
\end{xmpl}

\begin{xmpl}  
  We want to calculate an algebraic expression describing the set of
  infinite trees with the following properties: 1) every node is
  labeled by an element of a given set $E$, 2) every node has a finite
  (possibly empty) list of sons.  According to experience, this set
  could be expressed as the greatest solution of the equation
  \begin{eqnarray*}
    X & = & E \times X^{*}\,,
  \end{eqnarray*}
  that is, the final coalgebra of the functorial expression on the
  right.  On the other hand, we know that $X^{*}$ is the least
  solution of
  \begin{eqnarray*}
    Y & = & 1 + (Y \times X)\,,
  \end{eqnarray*}
  hence we guess that the desired algebraic expression is given by the
  $\mu$-term $\nu_{x}. (E \land \mu_{y}.  (\top \lor (y \land x ))$.
  We can verify that this guess is correct by transforming the
  $\mu$-term into a pointed parity game, according to proposition
  \ref{prop:termstogames}, the result being the game on the right of
  figure \ref{fig:unica}.
  \begin{figure}[h]
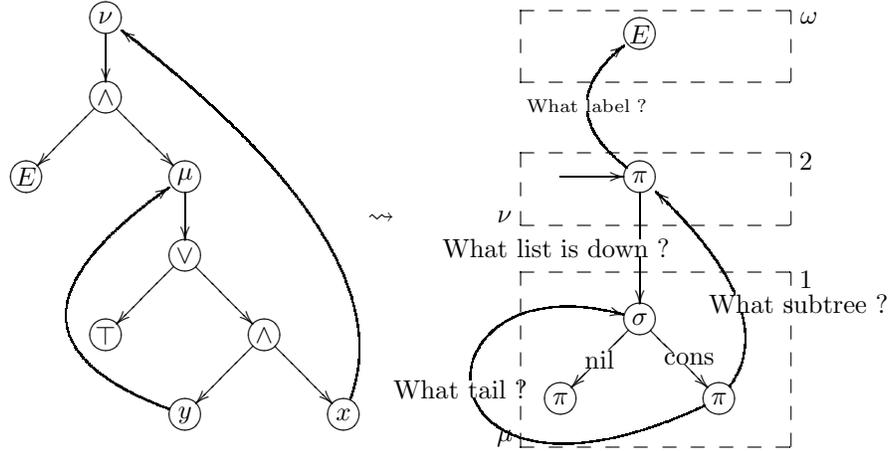

    $$
    \mygame[3em]{
      []!E{\nu}
      :[d]!E{\land}
      (:[ld]!E{E})
      :[rd]!E{\mu}
      :[d]!E{\vee}
      (
      [rrr]="I2",
      :[ld]!E{\top}
      [ll]="I1",,
      :[rd]!E{\land}
      (
      :[ld]!E{y}:@`{"I1"}"\mu",
      :[rd]!E{x}:@`{"I2"}"\nu"
      )
      )
    }
    \hspace{-0.3cm}
    \leadsto
    \hspace{0.0cm}
    \mygame[3em]{
      []
      :[r]!E{\pi}("\pi"="\nu")
      (:@/^2em/[u(1.8)]!E{E}|{\textrm{What label ?}})
      :[d(1.8)]!E{\sigma}="A"|{\makebox[8mm][r]{\textrm{What list is down ?}}}
      ("\sigma"="\mu",
      [rr]="I2", 
      :[ld]!E{\pi}="B"
      |{\makebox[4mm][l]{\nil}}
      ([ddll]="I1",[uull]="II1"), 
      :[rd]!E{\pi}="C"
      |{\makebox[4mm][l]{\cons}}(
      :@`{"I1","II1"}"\mu"|{
        \makebox[15mm][r]{\textrm{What tail ?}}
      }, 
      :@`{"I2"}"\nu"|{\makebox[8mm][l]{\textrm{What subtree ?}}}
      ) 
      )
      "E"
      [l(1.5)u(0.3)]("E"[r(1.5)d(0.2)]!L{}{\omega})
      "\nu"
      [l(1.5)u(0.3)]("\nu"[r(1.5)d(0.2)]!L{\nu}{2})
      "A"
      [l(1.5)u(0.6)]("C"[r(0.5)d(0.2)]!L{\mu}{1})
    }
    $$\vspace{-10mm}
    \caption{Infinite trees as winning strategies.}
    \label{fig:unica}
  \end{figure}
  It is possible to convince ourself that a labeled tree with those
  properties gives rise to a deterministic winning strategy for player
  $\sigma$ by interpreting a move by $\pi$ as a question about the
  tree. Conversely, every such strategy comes from a unique tree of
  this kind.
  
  It is worth examining infinite paths in this game.  Player $\sigma$
  cannot answer that a node has an infinite list of sons: this would
  be done by answering infinitely often ``cons'' to the question
  ``What tail ?'', without being asked the question ``What list is
  down ?''. The region visited infinitely often of maximal height  in
  such a play is colored by $\mu$, hence it is a loss for player
  $\sigma$.  On the other hand, player $\sigma$ can answer infinitely
  often ``cons'' provided the play is going down in examining the
  tree, that is, provided this answer is alternating with the question
  ``What list is down ?''. The maximal region visited infinitely often
  in such a play is colored by $\nu$, hence it is a win for player
  $\sigma$.
\end{xmpl}

\begin{xmpl}
  It is well known that infinite finitely branching trees can be
  encoded as infinite binary trees. Proposition \ref{prop:strategies}
  can be taken to be a generalization of this fact, in that it shows
  that the elements of every nullary parity functor can be encoded as
  infinite
  trees with a bounded out-degree.  
\end{xmpl}

\begin{xmpl}
  Charity \cite{MR96c:18005} is a programming language designed out of
  categorical principles, thus recursion and corecursion are in this
  context synonymous for the universal properties of initial and final
  coalgebras. An important principle of this programming language
  states that it is possible to define an arrow $f:\mu_{x}.T(x) \times
  B \rTo C$ from an algebra in context $g:T(C) \times B \rTo C$,
  provided $T$ is a strong categorical datatype \cite{MR94a:18008}.
  This means that $T$ comes with a natural transformation (a strength)
  $$
  \theta^{T}_{A,B} : T(A) \times B \rTo T(A \times B)
  $$
  satisfying associativity and unitary constrains.  The explicit
  characterization of set-theoretic parity functors allows the direct
  computation of a strength. If $A$ and $B$ are two collections of
  sets indexed by $P_{\omega}$, then we can associate to a strategy
  $\langle T,\lambda\rangle \in \mathcal{S}_{G,p}(A)$ and to a
  collection $b = \{ b_{x} \in B_{x}\}_{x \in P_{\omega}}$ the
  strategy $\langle T,\lambda^{b}\rangle \in \mathcal{S}_{G,p}(A
  \times B)$, where if $\gamma \in T$ and $\cod \gamma \in P_{\omega}$
  then $\lambda^{b}(\cod \gamma) = (\lambda(\cod \gamma),b_{\cod
    \gamma})$.
\end{xmpl}

%% file: conclusions.tex
The main result of this paper is the combinatorial characterization of
the functors on the category of sets and functions that are definable
by means of $\mu$-terms. This characterization leads to show that the
algebra of $\mu$-bicomplete categories, when realized in the category
of sets, is closely related to the theory of automata recognizing
infinite objects. For example an automaton recognizing -- by parity
condition -- infinite strings over the finite alphabet $\Sigma$ can be
described as a triple $\langle G,p,f \rangle$, where $\langle G,p
\rangle$ is a pointed parity game such that $P_{\omega} = \emptyset$,
$\epsilon(p) = \sigma$ for all $p \in P$, and $f : \sval{G}_{p} \rTo
\val{\nu_{x}.\bigvee_{\Sigma} x}$ is a function arising from labeling
the transitions of $G$ by symbols in $\Sigma$, function which turns
out to be definable in the language of $\mu$-bicomplete categories. A
subset $L \subseteq \Sigma^{\nnumbers}$ is recognizable if and only if
there exists such a triple $\langle G,p,f \rangle$, so that $L$ is the
image of $f$.  A main motivation for developing this work was indeed
to make available to this theory an algebraic language (the one of
$\mu$-bicomplete categories) which is alternative but also analogous
to the one of $\mu$-calculi \cite{AN01}.

The combinatorial characterization suggests also a way for enlarging
the collection of categories which are known to be $\mu$-bicomplete.
There are several toposes that occur in computer science -- for
example, the effective topos \cite{hyland} -- which are not complete
or cocomplete, in particular they are neither locally presentable nor
dually locally presentable.  A detailed analysis of the work presented
here could show that the explicit characterization of parity functors
can be carried within intuitionistic logic. If this were the case, the
characterization could be used to show that elementary toposes with a
natural number object are $\mu$-bicomplete.

Finally, it is an open problem to understand whether this
game-theoretic characterization is useful to understand $\mu$-functors
in arbitrary categories.  It is in general easier to understand
several algebraic equivalences in terms of game equivalences. We have
avoided to make precise this notion, but we conjecture that this can
be done so that two parity games are game-theoretic equivalent if and
only if their interpretations as functors are naturally isomorphic in
every $\mu$-bicomplete category.